\documentclass[prd,twocolumn,floatfix,nopacs,preprintnumbers,nofootinbib]{revtex4}

\usepackage{mathrsfs, amssymb, amsmath, mathtools, graphicx, xcolor, lastpage, footnpag, fancyhdr, bm, pbox, braket, bbold, appendix}
\usepackage[breaklinks,colorlinks,urlcolor=blue,citecolor=citcolor,linkcolor=lcolor]{hyperref}
\usepackage[utf8x]{inputenc}
\usepackage[caption = false]{subfig}
\usepackage[normalem]{ulem}

\definecolor{lcolor}{rgb}{0.5,0,0}
\definecolor{citcolor}{rgb}{0,0.3,0.0}

\newlength{\mycol}
\newcommand{\asbar}{\bar{\alpha}_{\mathrm{s}}}

\newcommand{\nf}{n_\mathrm{f}}
\newcommand{\as}{\alpha_\mathrm{s}}
\newcommand{\lqcd}{\Lambda_\text{QCD}}

\newcommand{\gev}{\ \textrm{GeV}}

\newcommand{\qs}{Q_\mathrm{s}}

\newcommand{\eq}{Eq.~}

\newcommand{\der}{\mathrm{d}}

\newcommand{\qso}{Q_\mathrm{s0}}

\DeclareMathAlphabet{\ib}{OML}{cmm}{b}{it}

\newcommand{\diag}[2][{}]{\pbox{\textwidth}{\includegraphics[scale=0.35]{{#2}}}}
\newcommand{\dia}[2][{}]{\pbox{\textwidth}{\includegraphics[scale=0.6]{{#2}}}}

\newcommand{\tr}[1]{\mathrm{tr}\left({{#1}}\right)}
\newcommand{\g}[1]{\mathcal{G}_{\ib{{#1}}}}

\newcommand{\uu}[1]{U_{\ib{{#1}}}}
\newcommand{\ud}[1]{U^\dagger_{\ib{{#1}}}}
\newcommand{\ld}[2]{L^{{#1}}_{\ib{{{#2}}}}}

\newcommand{\xt}{\ib{x}}

\newcommand{\nc}{N_\mathrm{c}}
\newcommand{\cd}{C_\mathrm{d}}
\newcommand{\cf}{C_\mathrm{F}}

\newcommand{\da}{d_\mathrm{A}}

\newcommand{\limonemod}{{\substack{\ib{w} \to \ib{z} \\ \ib{v} \to \ib{z'}}}}

\newcommand{\op}[2]{S^{({{#1}})}_{\ib{{{#2}}}}}

\newcommand{\lnc}{large-$N_\mathrm{c}$ }
\newcommand{\fnc}{finite-$N_\mathrm{c}$ }
\newcommand{\mata}[1]{\mathcal{A}_{{{#1}}}}
\newcommand{\matm}[1]{\mathcal{M}_{{{#1}}}}

\begin{document}
\graphicspath{{Images/}}

\author{T. Lappi}
\author{H. Mäntysaari}
\author{A. Ramnath}
\affiliation{
Department of Physics, University of Jyv\"askyl\"a %
 P.O. Box 35, 40014 University of Jyv\"askyl\"a, Finland
}
\affiliation{
Helsinki Institute of Physics, P.O. Box 64, 00014 University of Helsinki, Finland
}

\title{
Next-to-leading order  Balitsky-Kovchegov equation beyond large $\nc$
}


\preprint{}


\begin{abstract}
We calculate \fnc corrections to the next-to-leading order (NLO) Balitsky-Kovchegov (BK) equation. We find analytical expressions for the necessary correlators of six Wilson lines in terms of the two-point function using the Gaussian approximation. In a suitable basis, the problem reduces from the diagonalization of a six-by-six matrix to the diagonalization of a three-by-three matrix, which can easily be done analytically. We study numerically the effects of these \fnc corrections on the NLO BK equation. In general, we find that the \fnc corrections are smaller than the expected $1/\nc^2 \sim 10\%$. The corrections may be large for individual correlators, but have less of an influence on the shape of the amplitude as a function of the dipole size. They have an even smaller effect on the evolution speed as a function of rapidity.
\end{abstract}

\maketitle


\section{Introduction}

In hadronic collisions at high energies, large gluon densities are created by the emission of soft gluons carrying a small fraction of the longitudinal momentum of the parent~\cite{Abramowicz:2015mha}.
Nonlinear dynamics of gluons becomes important in such an environment, where parton densities eventually grow to become on the order of the inverse of the  QCD coupling $\as$. 
To describe QCD in this region, the Color Glass Condensate (CGC) effective field theory~\cite{Gelis:2010nm} has been developed.

In the CGC framework, cross sections for various scattering processes can be expressed in terms of correlators of Wilson lines. 
A Wilson line describes the eikonal propagation of a parton in the strong color field of the target. 
The energy dependence of the target color fields, and thus cross sections, is obtained by solving the so-called Jalilian-Marian--Iancu--McLerran--Weigert--Leonidov--Kovner (JIMWLK) equation~\cite{JalilianMarian:1996xn,JalilianMarian:1997jx,JalilianMarian:1997gr,Iancu:2001md}. 
This is a perturbative evolution equation that describes the Bjorken-$x$ dependence of a Wilson line. 
In phenomenological applications, it is usually convenient to work directly in terms of the Wilson line correlators, and to solve instead the Balitsky-Kovchegov (BK) equation~\cite{Kovchegov:1999yj,Balitsky:1995ub} for the dipole operator (correlator of two Wilson lines), which can be obtained from the JIMWLK equation in the \lnc limit. 

The CGC framework has been used extensively in phenomenological applications at leading order (LO) in $\as$, with the evolution equations resumming contributions $\sim \as \ln 1/x$ to all orders. 
Running coupling effects derived in Refs.~\cite{Kovchegov:2006vj,Gardi:2006rp,Albacete:2007yr,Balitsky:2006wa} (see also \cite{Lappi:2012vw}) can also be taken into account. 
The non-perturbative initial condition for the small-$x$ evolution is obtained by performing fits to the HERA structure function data~\cite{Abramowicz:2015mha,H1:2018flt}, for example in Refs.~\cite{Albacete:2010sy,Albacete:2012rx,Lappi:2013zma,Mantysaari:2018zdd} (see also \cite{Rezaeian:2012ji,Mantysaari:2018nng}). 
The obtained initial condition can then be used for various calculations, for example particle production in proton-nucleus collisions~\cite{Tribedy:2011aa,Goncalves:2012bn,Lappi:2013zma,Ducloue:2015gfa,Ducloue:2016pqr,Albacete:2016tjq,Ducloue:2017kkq,Mantysaari:2019nnt}. 
In the future, the nuclear deep inelastic scattering (DIS) experiments at the Electron Ion Collider (EIC)~\cite{Aschenauer:2017jsk,Accardi:2012qut} in the US, at the LHeC~\cite{AbelleiraFernandez:2012cc} at CERN and at the EicC in China~\cite{Chen:2018wyz} will provide a vast amount of precise data from clean DIS processes. 
These experiments will be able to probe the nuclear structure where nonlinearities are enhanced by roughly $A^{1/3}$ higher densities compared to the proton. 
Before the EIC, similar studies limited to the photoproduction region can be performed in ultra-peripheral heavy-ion collisions~\cite{Bertulani:2005ru,Klein:2019qfb}.

In order to quantitatively study nonlinear dynamics in high-energy scattering processes (and especially at the future EIC), it is crucial to move beyond LO accuracy. 
The next-to-leading order (NLO) evolution equations are available: the NLO BK equation was derived in Ref.~\cite{Balitsky:2008zza} and the NLO JIMWLK equation was derived in Refs.~\cite{Balitsky:2013fea,Kovner:2013ona}. 
Similarly, the impact factors are becoming available at NLO for some processes: inclusive DIS~\cite{Ducloue:2017ftk,Beuf:2017bpd,Beuf:2016wdz,Balitsky:2010ze,Hanninen:2017ddy} (in the case of massless quarks), exclusive vector meson production~\cite{Boussarie:2016bkq,Escobedo:2019bxn} (see also~\cite{Lappi:2020ufv}) and particle production in proton--nucleus collisions~\cite{Chirilli:2012jd}. 
However, the phenomenological applications of these are still developing~\cite{Ducloue:2017dit,Ducloue:2016shw,Watanabe:2015tja,Altinoluk:2014eka,Stasto:2013cha,Liu:2019iml}. 

The BK equation is usually solved in the \lnc limit. In the LO case, the \lnc limit makes it possible to express the four-point correlator of fundamental representation Wilson lines in terms of the two-point function. 
In detailed numerical studies, it has been shown that the \fnc corrections are smaller than the naive expectation of $\mathcal{O}(1/\nc^2)$~\cite{Rummukainen:2003ns,Kovchegov:2008mk}. 
At NLO, the equation involves six-point functions of fundamental Wilson lines that must similarly be expressed in terms of the two-point function in order to close the equation.
The purpose of this work is to see if the \fnc corrections are similarly small in the case of the NLO equation, where all corrections of the order $\as^2$ are taken into account.

In order to numerically solve the BK equation at finite~$\nc$, we use the Gaussian approximation~\cite{Fujii:2006ab,Marquet:2010cf,Dominguez:2011wm}
to derive analytical parametric equations for the six-point correlators in terms of the two-point correlators. 
We study numerically the \fnc corrections to the these correlators, and also their effect on the NLO BK evolution. 
In addition to the BK equation, higher-point correlators are needed in the calculations of multi-particle correlations in the CGC framework, see eg. Refs.~\cite{Marquet:2007vb,Lappi:2012nh,Dusling:2017aot,Dominguez:2012ad}. 

The structure of the paper is as follows. 
In Section II, we introduce the NLO BK equation and provide both the \lnc and \fnc expressions for the correlators that will be studied. 
In Section III, we introduce the Gaussian approximation, explain the diagrammatic notation used in the rest of the paper and then explain the analytical calculation done for finding the parametric equations for the six-point correlators. 
Section IV contains the numerical results obtained from using the analytical expressions for the six-point correlators to solve the BK equation. 
Finally, we end with a few concluding remarks and a summary of our main results. 


\section{The BK equation at NLO}

For any product of $n/2$ pairs of fundamental Wilson lines $\uu{} \ud{}$, we use the notation 
\begin{align}
S^{(n)}_{\ib{x}_1, \ib{x}_2, \ldots , \ib{x}_{n-1}, \ib{x}_n} := \frac{1}{\nc} \tr{\uu{x_\mathrm{1}} \ud{x_\mathrm{2}} \ldots \uu{x_\mathrm{n-1}} U^\dagger_{\ib{x}_n}}.
\end{align}
The NLO BK equation in the case of zero active quark flavors
($n_f = 0$) reads~\cite{Balitsky:2008zza}
\begin{multline}
\label{eq:nlobk}
\partial_Y \left\langle \op{2}{x,y} \right\rangle = \frac{\as \nc}{2\pi^2} K_1^\text{BC} \otimes \langle D_1 \rangle 
\\
+ \frac{\as^2 \nc^2}{16\pi^4} K_{2,1} \otimes \langle D_{2,1} \rangle +  \frac{\as^2 \nc^2}{16\pi^4} K_{2,2} \otimes \langle D_{2,2} \rangle 
\\
+ \mathcal{O}(n_f),
\end{multline}
where the brackets $\langle \rangle$ refer to the expectation value over  target color field configurations. 
The kernels are
\begin{widetext}
\begin{align}
\label{eq:orig_k1}
K^\textnormal{BC}_1 =& \; \frac{r^2}{X^2Y^2} \left[ 1+\frac{\as\nc }{4\pi} \left( \frac{\beta}{\nc} \ln r^2\mu^2 - \frac{\beta}{\nc} \frac{X^2-Y^2}{r^2} \ln \frac{X^2}{Y^2} + \frac{67}{9} - \frac{\pi^2}{3} - \frac{10}{9} \frac{\nf}{\nc} - 2\ln \frac{X^2}{r^2} \ln \frac{Y^2}{r^2} \right) \right],
\\
K_{2,1} =& \; -\frac{4}{Z^4} + \left\{ 2\frac{X^2 Y'^2 + X'^2Y^2 - 4r^2 Z^2}{Z^4(X^2Y'^2 - X'^2Y^2)} + \frac{r^4}{X^2Y'^2 - X'^2Y^2}\left[ \frac{1}{X^2Y'^2} + \frac{1}{Y^2X'^2} \right] \right. \nonumber
\\
& \qquad \qquad \qquad \qquad \qquad \qquad \qquad \qquad \qquad \qquad \qquad \qquad + \left. \frac{r^2}{Z^2}\left[ \frac{1}{X^2Y'2} - \frac{1}{X'^2Y^2} \right] \right\}  
\times \ln \frac{X^2Y'^2}{X'^2Y^2}, 
\\
K_{2,2} =& \; \left\{ \frac{r^2}{Z^2} \left[ \frac{1}{X^2Y'^2} + \frac{1}{Y^2X'^2}  \right]  - \frac{r^4}{X^2Y'^2X'^2Y^2}  \right\} \ln \frac{X^2Y'^2}{X'^2Y^2}.
\end{align}
\end{widetext}

The convolutions $\otimes$ in \eq\eqref{eq:nlobk} denote integrations over the transverse coordinate $\ib{z}$ (in $K_1^\text{BC}$) or $\ib{z}$ and $\ib{z'}$ (in $K_{2,1}$ and $K_{2,2}$). 
We use the notation 
$ r^2 = (\ib{x}-\ib{y})^2, 
X^2 = (\ib{x} - \ib{z})^2, 
X'^2 = (\ib{x} - \ib{z'})^2,
Y^2 = (\ib{y} - \ib{z})^2,
Y'^2 = (\ib{y} - \ib{z'})^2$
and $Z^2 = (\ib{z} - \ib{z'})^2$.
We note that the kernel proportional to $n_f$ is also available~\cite{Balitsky:2008zza}. 
Since the purpose of this work is to study the importance of the \fnc corrections in the NLO BK equation, we do not include contributions proportional to $n_f$. 
The \fnc effects could be expected to be  similar to the $n_f = 0$ case.

The Wilson line operators appearing in \eq\eqref{eq:nlobk} are
\begin{align}
\label{eq:d1}
\langle D_1 \rangle =& \left\langle \op{2}{x,z} \op{2}{z,y} \right\rangle - \left\langle \op{2}{x,y} \right\rangle, 
\\
\label{eq:d21}
\langle D_{2,1} \rangle =& \left\langle \op{2}{x,z} \op{2}{z,z'} \op{2}{z',y} \right\rangle -
\frac{1}{\nc^2} \left\langle \op{6}{x,z,z',y,z,z'} \right\rangle \nonumber
\\
& \qquad \qquad \qquad \qquad \qquad \qquad - (\ib{z'} \to \ib{z}), 
\\
\label{eq:d22}
\langle D_{2,2} \rangle =& \left\langle \op{2}{x,z} \op{2}{z,z'} \op{2}{z',y} \right\rangle - (\ib{z'} \to \ib{z}). 
\end{align}
Although the original NLO BK equation in the form presented in Ref.~\cite{Balitsky:2008zza} does not contain the subtraction $\ib{z'} \to \ib{z}$ in $D_{2,2}$, we have introduced the subtraction to improve numerical stability. This subtraction term has no effect on the final evolution because the integral of $K_{2,2}$ over $\ib{z}'$ vanishes if the Wilson line operator term does not depend on $\ib{z}'$ (see Ref.~\cite{Balitsky:2008zza}).
We will refer throughout this work to two pieces of the right side of Eq.~\eqref{eq:nlobk} as the
\begin{itemize}
\item $\frac{\as \nc}{2\pi^2} K_1^\text{BC} \otimes \langle D_1 \rangle \sim$ ``LO-like'' contribution,
\item $\frac{\as^2 \nc^2}{16\pi^4} K_{2,1} \otimes \langle D_{2,1} \rangle +  \frac{\as^2 \nc^2}{16\pi^4} K_{2,2} \otimes \langle D_{2,2} \rangle \\ \sim $ ``NLO-like'' contribution.
\end{itemize}
In other words, we separate the terms in the NLO BK equation by the types of Wilson line correlators, not by the order in $\as$. 
Thus, the LO-like contribution also includes a significant $\as^2$ correction.

The interpretation of the NLO BK equation is that one considers all possible ways to emit either one or two gluons, at transverse coordinates $\ib{z}$ and $\ib{z}'$, from the boosted dipole consisting of quarks at transverse coordinates $\ib{x}$ and $\ib{y}$. 
The effect of the boost is that instead of the original dipole projectile, the original quarks and the emitted gluons scatter off the target color field. 
As such, the evolution can be seen to describe the evolution of the projectile probing the target structure. 
On the other hand, the emitted gluons can also be taken to be a part of the target wave function, in which case the boost corresponds to the evolving target color field as probed by the original projectile. 
For a more detailed discussion on the NLO evolution in the projectile or target wave function, the reader is referred to Ref.~\cite{Ducloue:2019ezk}.

The NLO BK equation is known to be unstable~\cite{Lappi:2015fma} due to the large contributions enhanced by the large double transverse logarithm $\ln X^2/r^2 \ln Y^2/r^2$. 
We resum these contributions to all orders following the procedure developed in Ref.~\cite{Iancu:2015vea}, which was numerically confirmed in Ref.~\cite{Lappi:2016fmu} to result in a stable evolution  (see also Ref.~\cite{Beuf:2014uia} for an equivalent resummation of the same double logarithms). 
In addition, we include the running of the QCD coupling by noticing that the terms proportional to the beta function coefficient $\beta$ in Eq.~\eqref{eq:orig_k1} should be resummed into the running coupling. 
We implement this resummation by following the Balitsky prescription from Ref.~\cite{Balitsky:2006wa}. 
Both running coupling and double transverse logarithm resummations are included by modifying the kernel $K_1^\text{BC}$ as
\begin{multline}
\label{eq:full_k1}
\frac{\as \nc}{2\pi^2} K_1^\text{BC} \to \frac{\as(r) \nc}{2\pi^2} K_\text{DLA} 
\\
\times
\left[\frac{r^2}{X^2Y^2} + \frac{1}{X^2} \left(\frac{\as(X)}{\as(Y)}-1\right) + \frac{1}{Y^2} \left(\frac{\as(Y)}{\as(X)}-1\right) \right] 
\\
+ K_1^\text{fin}.
\end{multline}
The double log corrections to all orders are taken into account by the factor
\begin{align}
K_\text{DLA} 
= \frac{J_1\left(2\sqrt{\asbar x^2}\right)}{\sqrt{\asbar x^2}}, 
\end{align}
where $\asbar = \as \nc/\pi$.
The double logarithm here is $x = \sqrt{\ln X^2/r^2 \ln Y^2/r^2}$. 
If $\ln X^2/r^2 \ln Y^2/r^2 < 0$, then an absolute value is used and the Bessel function is changed from $J_1 \to I_1$ (see Ref.~\cite{Iancu:2015vea}).
The scale of the coupling in $K_\text{DLA}$ is determined by the smallest dipole $\min\{r^2,X^2,Y^2 \}$.

In addition to the double log contributions, one can also resum a set of higher-order contributions enhanced by single transverse logarithms, as shown in Ref.~\cite{Iancu:2015joa}. 
For the purposes of this paper, this resummation is not necessary and is excluded for simplicity.
In this running coupling prescription, we keep the order $\as^2$ terms in the kernel $K_1^\text{BC}$ that are not proportional to the beta function. 
These are included in the term $K_1^\text{fin}$, which reads
\begin{align}
K_1^\text{fin} = \frac{\as^2(r) \nc^2}{8\pi^3} \frac{r^2}{X^2Y^2} \left( \frac{67}{9} - \frac{\pi^2}{3} - \frac{10}{9} \frac{\nf}{\nc} \right).
\end{align}
The strong coupling constant in the transverse coordinate space is evaluated as
\begin{align}
\as(r) = \frac{4\pi}{\beta \ln\left\{\left[ \left( \frac{\mu_0^2}{\lqcd^2} \right)^{\frac{1}{c}}
+ \left( \frac{4C^2}{r^2 \lqcd^2} \right)^{\frac{1}{c}} \right]^{c} \right\}},
\end{align}
where $\beta = (11\nc - 2n_f)/\nc$.
We use\footnote{A generic estimate~\cite{Kovchegov:2006vj,Lappi:2012vw} would be $C^2=e^{-2\gamma_E} \approx 0.32$. We use a larger value $C^2=1$ which results in slightly slower evolution, as the $C^2$ is usually taken to be a free parameter controlling the coordinate space scale.}  $C^2=1$ and $\mu_0/\lqcd=2.5$ in our numerical calculations, which freezes the coupling at $\as(r\to\infty)=0.762$ in the infrared, and $c=0.2$ which controls the transition to the infrared region. 

The initial condition for the BK equation is taken from the McLerran-Venugopalan (MV) model~\cite{McLerran:1993ni,McLerran:1997fk}. 
In the MV model, the color charge density is assumed to be a random Gaussian variable, with a zero expectation value and a variance proportional to the local saturation scale $Q_s^2$. 
The dipole correlator in the MV model is written as
\begin{align}
\label{eq:mv}
\left\langle \op{2}{x,y} \right\rangle_\mathrm{MV} = \exp \left[ -\frac{r^2 \qso^2}{4} \ln \left(\frac{1}{r\lqcd} + e \right) \right].
\end{align}
Here, the constant $e$ acts as an infrared regulator. 
We use $\lqcd=0.241\gev$ and $\qso^2=1\gev^2$ in the numerical analysis.
In analytical studies of the correlators of Wilson lines in specific ``line'' coordinate configurations in Sec.~\ref{sec:line}, we use the GBW~\cite{GolecBiernat:1998js} form for the dipole correlator
\begin{align}
\label{eq:gbw}
\left\langle \op{2}{x,y} \right\rangle_\mathrm{GBW} = \exp \left( - \frac{r^2 \qso^2}{4}  \right).
\end{align}
In principle, the resummation procedure for the double transverse logs would also change the initial condition, as discussed in Refs.~\cite{Iancu:2015vea,Lappi:2016fmu}. 
However, since the initial condition is a non-perturbative input for the evolution, we consider Eq.~\eqref{eq:mv} to be the non-perturbative initial condition for the resummed evolution as well. 
For the purposes of this paper, the actual form of the initial condition is not relevant.

We compare the \fnc version of the BK equation presented above to the equation obtained in the \lnc limit, which has been studied numerically in Refs.~\cite{Lappi:2015fma,Lappi:2016fmu}. 
In this limit, one can drop operators suppressed by $1/\nc$ and the correlators in Eq.~\eqref{eq:nlobk} become
\begin{align}
\label{eq:largenc-d1}
\langle D_1 \rangle \xrightarrow{\nc \to \infty} 
\langle D_1 \rangle_{\nc \to \infty} 
= \left\langle \op{2}{x,z} \right\rangle \left\langle \op{2}{z,y} \right\rangle - \left\langle \op{2}{x,y} \right\rangle,
\end{align}
\begin{align}
\label{eq:largenc-d2}
\langle D_{2,1} \rangle & \xrightarrow{\nc \to \infty } \langle D_2 \rangle_{\nc \to \infty}, 
\\
\langle D_{2,2} \rangle  & \xrightarrow{\nc \to \infty } \langle D_2 \rangle_{\nc \to \infty}, 
\end{align}
with 
\begin{equation}
\langle D_2 \rangle_{\nc \to \infty} =     
\left\langle \op{2}{x,z} \right\rangle \left\langle \op{2}{z,z'} \right\rangle \left\langle \op{2}{z',y} \right\rangle 
- \left\langle \op{2}{x,z} \right\rangle \left\langle \op{2}{z,y} \right\rangle.
\end{equation}



\section{Six-point functions in the Gaussian approximation}

\subsection{The Gaussian approximation}

At large $\nc$, all Wilson line operators present in the NLO BK equation can be expressed solely in terms of dipole correlators, as can be seen from Eqs.~\eqref{eq:largenc-d1} and \eqref{eq:largenc-d2}. 
At finite $\nc$, on the other hand, the higher-point functions $\left\langle \op{2}{x,z} \op{2}{z,y} \right\rangle$, $\left\langle \op{2}{x,z} \op{2}{z,z'} \op{2}{z',y} \right\rangle$ and $\left\langle \op{6}{x,z,z',y,z,z'} \right\rangle$ are needed.
An expression for the four-point function in terms of two-point functions has been derived using the Gaussian approximation (see e.g.~\cite{Marquet:2010cf}).
This makes it possible to obtain a closed form for the LO BK equation at finite $\nc$.
The purpose of this work is to compute also the six-point functions using the Gaussian approximation, in order to obtain a closed \fnc BK equation at NLO accuracy.

In the Gaussian approximation, all correlators are parametrized by a single two-point function, and all higher-point functions can then be expressed in terms of this function. 
The initial condition for the small-$x$ evolution is usually assumed to be Gaussian (e.g. as in the MV model), but it is not clear \emph{a priori} that the Gaussian approximation is valid after the evolution. However, numerical studies of the JIMWLK equation~\cite{Dumitru:2011vk,Lappi:2015vta} have not found any indication of major effects breaking the validity of this approximation.

We use the diagrammatic notation of Refs.~\cite{Kovchegov:2008mk,Marquet:2010cf,Lappi:2016gqe}, in which Wilson lines are denoted as
\begin{align}
\uu{x} = \; \dia{emp1} \dia{tar1} \dia{emp1} \; ,
\\
\ud{x} = \; \dia{emp1} \dia{tar1cc} \dia{emp1} \; .
\end{align}
The projectile transverse coordinate is $\xt$, the lightcone time axis runs from right to left and the blue vertical line represents the target background field.
In the Gaussian approximation, the correlator for some Wilson line operator $\mathcal{O} [U]$ is approximated as an integral over a \emph{parametrization rapidity} $\eta$ of a single two-point correlator $G_{\ib{u}_1,\ib{u}_2}$:
\begin{multline}
\label{eq:gt}
\left\langle \mathcal{O}[U] \right\rangle_\eta = 
\\
\exp \left\{ - \frac{1}{2} \int^\eta \der \tilde{\eta} \int_{\ib{u}_1,\ib{u}_2} G_{\ib{u}_1,\ib{u}_2} (\tilde{\eta}) \ld{a}{u_\textrm{1}} \ld{a}{u_\textrm{2}} \right\} \mathcal{O}[U].
\end{multline} 
The transverse integrals are denoted as $\int_\ib{u} = \int \der^2 \ib{u}$ and $\ld{a}{u}$ is a Lie derivative that acts on Wilson lines according to
\begin{align}
\ld{a}{u} \uu{x} 
&= -ig \delta^{(2)}(\ib{x}-\ib{u}) t^a \uu{x} 
\\
&= - i g \; \dia{emp1} \dia{tar1} \dia{glu1} \; .
\end{align}
The structure as an exponential of a two-point function (as in the MV model \cite{McLerran:1993ni}) is what makes this a ``Gaussian'' approximation.
In practice, for gauge invariant (color singlet) operators, the two-point function $G_{\ib{u}_1,\ib{u}_2}$ always appears in the linear combination
\begin{multline}
\g{u_\textrm{1},u_\textrm{2}} (\eta) := \int^\eta \der \tilde{\eta} \Big[ G_{\ib{u}_1,\ib{u}_2} (\tilde{\eta})
\\
- \frac{1}{2} \Big( G_{\ib{u}_1,\ib{u}_1} (\tilde{\eta}) + G_{\ib{u}_2,\ib{u}_2} (\tilde{\eta}) \Big) \Big].
\end{multline} 
For the integrand, we use the notation $\mathcal{G}' \coloneqq \partial_\eta \; \mathcal{G}$. Physical observables only depend on the integrated $\g{}$ and not on the integrand $\mathcal{G}'$. Thus, for the purpose of our calculation, where we need to relate higher-point functions of Wilson lines to the two-point function, there is some freedom in choosing the parametrization rapidity. We use this freedom in such a way that the $\eta$ and transverse coordinate dependences of $\mathcal{G}'$ factorize\footnote{This assumption has the effect that the transition matrices $\matm{} (\eta)$ (introduced below in \eq\eqref{eq:gtder}) at different rapidities $\eta$ commute with each other. This turns the path ordered exponential of $\matm{} (\eta)$ into a normal exponential. This, in turn, makes it possible to relate higher-point functions to the two-point function without any further assumptions about the $\eta$ dependence of $\mathcal{G}'$. In the terminology of Ref.~\cite{Marquet:2010cf}, we use a ``rigid exponentiation'' instead of the ``Gaussian Truncation''. The ``Gaussian Truncation''  would imply equating the parametrization rapidity $\eta$ with the evolution rapidity $Y$. See the related discussion in Refs.~\cite{Iancu:2011ns,Lappi:2016gqe}.}, as is usually done when employing the Gaussian approximation (see e.g.~\cite{Blaizot:2004wu,Blaizot:2004wv,Fujii:2006ab,Dominguez:2008aa,Dominguez:2011wm,Dumitru:2011vk,Lappi:2012nh}) in phenomenological applications. 
We henceforth omit the explicit $\eta$ dependence of $\g{}$ for brevity. 

As an example of the procedure for finding the parametric equation for a correlator using Eq.~\eqref{eq:gt}, we illustrate the steps for the dipole operator $\left\langle \op{2}{x,y} \right\rangle$.
A more practical form of Eq.~\eqref{eq:gt} is to write the integral over the parametrization rapidity $\eta$ in differential form.
Then, we have
\begin{align}
\label{eq:gtdipole}
\partial_\eta \left\langle \op{2}{x,y} \right\rangle = - \mathcal{M}(\eta) \left\langle \op{2}{x,y} \right\rangle,
\end{align} 
where $\matm{}$ is the so-called transition matrix formed by operating with the argument of the exponential in Eq.~\eqref{eq:gt} on the operator of interest.
In this case, there are four contributions from acting with the Lie derivatives on a product of two Wilson lines:
\begin{align}
& \ld{a}{u_\textrm{1}} \ld{a}{u_\textrm{2}} \Big( \uu{x} \otimes \ud{y} \Big)
= \ld{a}{u_\textrm{1}} \ld{a}{u_\textrm{2}} \; \dia{emp2} \dia{tar2} \dia{emp2}
\\
\sim & \; \dia{emp2} \dia{tar2} \dia{gtuu} \;
+ \; \dia{emp2} \dia{tar2} \dia{gtud} \nonumber
\\
& + \; \dia{emp2} \dia{tar2} \dia{gtdu} \;
+ \; \dia{emp2} \dia{tar2} \dia{gtdd} \;.
\end{align} 
From this sum, we need to factorize out the original operator $\uu{x} \otimes \ud{y}$, so we use the Fierz identity 
\begin{align}
\label{eq:fierz}
2 \; \dia{fierz_tt} \; = \; \dia{fierz_2} \; - \; \frac{1}{\nc} \; \dia{fierz_1} \;.
\end{align} 
There is only one way to join the endpoints of the Wilson lines into a singlet operator. This is to trace over them, i.e. wedging them between $\frac{1}{\sqrt{\nc}} \; \dia{m1l2}$ and $\frac{1}{\sqrt{\nc}} \; \dia{m1r2} \;$.
Doing so and performing the remaining operations in the exponent of Eq.~\eqref{eq:gt}, we get
\begin{multline}
\partial_\eta \left\langle \op{2}{x,y} \right\rangle = - \cf \g{x,y}' \left\langle \op{2}{x,y} \right\rangle
\\
\implies \frac{1}{\nc} \; \diag{m1l2} \diag{tar2} \diag{m1r2} \;
= \left\langle \op{2}{x,y} \right\rangle = e^{- \cf \g{x,y}},
\label{eq:gtdip}
\end{multline} 
where $\cf = \frac{\nc^2 - 1}{2 \nc}$. 
Since the operators $\diag{m1l2}$ and $\diag{m1r2}$ were normalized, the initial condition at $\eta = 0$ for the differential equation \eqref{eq:gtdipole} is given by trivial Wilson lines equal to the identity matrix in the absence of a color field.
This is the well-known parametric equation for the dipole correlator~\cite{Marquet:2010cf} in the Gaussian approximation. 

In the case of $n$-point correlators larger than the dipole, the operator $\left\langle \mathcal{O}[U] \right\rangle$ in Eq.~\eqref{eq:gt} is actually an $n \times n$ matrix of correlators, denoted $\mata{} (\eta)$, and Eq.~\eqref{eq:gtdipole} becomes an $n \times n$ matrix differential equation
\begin{align}
\label{eq:gtder}
\partial_\eta \mata{} (\eta) = - \matm{} (\eta) \mata{} (\eta).
\end{align} 
By construction, $\matm{}$ is a symmetric matrix, so there are at most $\sum^n_{i=1} i = n(n+1)/2$ distinct elements, not $n \times n$. 

For example, a product of six Wilson lines is represented in this notation as
\begin{align}
\uu{z} \otimes \ud{z'} \otimes \uu{v} \otimes \ud{y} \otimes \uu{x} \otimes \ud{w} = \; \dia{tarlabel} \; .
\end{align}
(the haphazard assignment of coordinate labels is convenient for the NLO BK equation and will become clear when constructing the transition matrix).
The notation here means that this product is actually a matrix with six open indices on the left and another six on the right; we denote them as
\begin{align}
\dia{empljk} \dia{tar} \dia{empril} \; . \nonumber
\end{align}
Since there are six possible ways to join the endpoints of these Wilson lines to form a correlator, Eq.~\eqref{eq:gtder} is a six-by-six matrix differential equation, as opposed to the much simpler one-dimensional problem illustrated for the dipole correlator. 

In analogy to the procedure for the dipole operator, the procedure to use Eq.~\eqref{eq:gtder} to find parametric equations for the six-point correlators is as follows:
\begin{enumerate}
\item Choose a multiplet basis, represented as a column vector  $\ib{B}$,  of $n=6$ color structures for the space of all six-point correlators.
Each element will have six open color indices, which can contract with the open indices on the left of $\uu{z} \otimes \ud{z'}  \otimes\uu{v} \otimes \ud{y}  \otimes\uu{x} \otimes \ud{w}$.
For example, one choice for an element of $\ib{B}$ could be
\begin{align}
\frac{1}{\sqrt{\nc^{3}}} \; \dia{m1l} \dia{emprjk} \;
= \frac{1}{\sqrt{\nc^{3}}} \delta_{j_1, k_1} \delta_{j_2, k_2} \delta_{j_3, k_3}
\end{align}
and the corresponding element for the other end of the Wilson line is then
\begin{align}
\frac{1}{\sqrt{\nc^{3}}} \; \dia{emplil} \dia{m1r} \;
= \frac{1}{\sqrt{\nc^{3}}} \delta_{i_1, l_1} \delta_{i_2, l_2} \delta_{i_3, l_3}.
\end{align}
The prefactor is a normalization constant found by squaring the basis element.
\item Construct the correlator matrix $\mata{}$ by taking $\ib{B} \left( \uu{z} \otimes \ud{z'}  \otimes\uu{v} \otimes \ud{y}  \otimes\uu{x} \otimes \ud{w} \right) \ib{B}^\mathrm{T}$. The elements of $\ib{B}$ on the left are contracted with the open color indices on the left of the Wilson lines, and the elements of $\ib{B}^\mathrm{T}$ with the open indices on the right.
For example, using the basis element shown above, we have for one of the 36 elements in $\mata{}$,
\begin{multline}
\frac{1}{\nc^3} \; \dia{m1l} \dia{emprjk} \dia{tar} \dia{emplil} \dia{m1r}
\\ \\
= \frac{1}{\nc^3} \tr{\uu{z} \ud{z'}} \tr{\uu{v} \ud{y}} \tr{\uu{x} \ud{w}}.
\end{multline}
\item Construct the transition matrix $\matm{}$ by summing (for each element in $\mata{}$) all possible one-gluon diagrams obtained with the double Lie derivative operator and rewriting the result in terms of elements of $\mata{}$.
\item Solve \eq\eqref{eq:gtder} by exponentiating $\matm{}$ to find expressions for each element in $\mata{}$, using as an initial condition the correlator matrix $\mata{}$ corresponding to Wilson lines equal to the identity matrix.
\end{enumerate}


\subsection{Choosing a basis}

Starting from a product of six Wilson lines, there are six ways to form multiplets by joining endpoints in all possible ways:
\begin{align}
\diag{m1l}
\qquad
\diag{m2l}
\qquad
\diag{m3l}
\qquad
\diag{m4l}
\qquad
\diag{m5l}
\qquad
\diag{m6l} \; .
\end{align}
The simplest way to construct an orthonormal basis from these would be to use color algebra to choose
\begin{align}
\label{eq:basissimple}
\ib{B} \coloneqq
\begin{pmatrix} 
\frac{1}{\sqrt{\nc^{3}}} \; \diag{m1l}
\\ \\
\sqrt{\frac{4}{\nc \da}} \; \diag{m1la}
\\ \\
\sqrt{\frac{4}{\nc \da}} \; \diag{m1lb}
\\ \\
\sqrt{\frac{4}{\nc \da}} \; \diag{m1lc}
\\ \\
\frac{1}{i} \sqrt{\frac{8}{\nc \da}} \; \diag{fl}
\\ \\
\sqrt{\frac{8}{\cd \da}} \; \diag{dl}
\end{pmatrix} \; .
\end{align}
The blue lines denote gluons and the last two elements of $\ib{B}$ represent the anti-symmetric and symmetric structure constants, respectively, $f_{abc} = -2i \tr{[t_a,t_b], t_c}$ and $d_{abc} = 2 \tr{\{t_a, t_b\}, t_c}$. 
The color factors are $\da = \nc^2 - 1$ and $\cd = \frac{\nc^2 - 4}{\nc}$. 
The next step would be to use this basis to construct the correlator matrix $\mata{}$ and the transition matrix $\matm{}$.
However, doing so results in a matrix differential equation $\partial_\eta \mata{} (\eta) = - \matm{} (\eta) \mathcal{A} (\eta)$, whose complicated solution is the matrix exponential of a six-by-six matrix $\matm{}$. 

For our case, a better way to proceed is to exploit the structure of the six-point correlators that are actually needed for the NLO BK equation. 
Since there are only four distinct coordinates in these particular correlators, we make the coordinate assignments
\begin{align}
\dia{tarlabel}
\qquad \xrightarrow[\ib{v} \to \ib{z'}]{\ib{w} \to \ib{z}} \qquad
\dia{tarlabel2} \; .
\end{align}
It is easy to see from this that there is one way to join the endpoints such that  in the limit $\ib{v} \to \ib{z'}, \ib{w} \to \ib{z}$, four Wilson lines cancel (due to unitarity). 
The result simplifies to a single trace:
\begin{align}
\frac{1}{\sqrt{\nc^{3}}} \; \diag{m6l} \quad \diag{tar} \quad \diag{m6r} \; \frac{1}{\sqrt{\nc^{3}}}
\quad \xrightarrow[\ib{v} \to \ib{z'}]{\ib{w} \to \ib{z}} \quad
\frac{1}{\nc^3} \nc^2 \tr{\uu{x} \ud{y}}.
\end{align}
So choosing $\frac{1}{\sqrt{\nc^{3}}} \; \diag{m6l} \;$ as one of our basis elements allows one dimension of our six-dimensional space of operators to decouple, giving the equation for the dipole correlator. 
Similarly, the choice of two more particular basis elements results in two correlators that reduce to four-point functions; one due to the limit $\ib{v} \to \ib{z'}$ and the other due to the limit $\ib{w} \to \ib{z}$. Thus, we can expect to choose a further two basis elements such that two more dimensions decouple from the remaining five, corresponding to the equation for the four-point operators.
These two basis elements can be chosen as 
\begin{align}
\frac{1}{\sqrt{2 \nc \da}} & \left[ - \; \diag{m2l} \; + \; \diag{m3l} \; \right] \nonumber
\end{align}
and
\begin{align}
\frac{1}{\sqrt{2 \nc \da}} & \left[ - \; \diag{m2l} \; - \; \diag{m3l} \; + \frac{2}{\nc} \; \diag{m6l} \; \right]. \nonumber
\end{align}

We will choose the remaining three basis elements such that they are orthonormal to the three already chosen, resulting in the final basis vector
\begin{multline}
\label{eq:basis}
\tilde{\ib{B}} \coloneqq
\\
\begin{pmatrix} 
\frac{\sqrt{2}}{\nc \sqrt{\da \cd}}
\left[ \frac{\nc}{2} \diag{m1l} - \diag{m2l} - \diag{m3l} - \diag{m4l} + \frac{\nc}{2} \diag{m5l} + \frac{2}{\nc} \diag{m6l} \right]
\\
\frac{1}{\sqrt{2 \nc \da}} \left[ - \; \diag{m1l} \; + \; \diag{m5l} \; \right]
\\
\frac{1}{\sqrt{\nc \da}}
\left[ - \; \diag{m4l} \; + \frac{1}{\nc} \; \diag{m6l} \; \right]
\\
\frac{1}{\sqrt{2 \nc \da}} \left[ - \; \diag{m2l} \; + \; \diag{m3l} \; \right]
\\
\frac{1}{\sqrt{2 \nc \da}} \left[ - \; \diag{m2l} \; - \; \diag{m3l} \; + \frac{2}{\nc} \; \diag{m6l} \; \right]
\\
\frac{1}{\sqrt{\nc^{3}}} \;
\diag{m6l}
\end{pmatrix}.
\end{multline}
Since this basis is orthonormal, the correlator matrix at the initial condition $\mata{} (\eta=0) $ will just be the identity matrix.


\subsection{Constructing the correlator matrix and the transition matrix}
\label{sec:matm}

Due to this choice of basis $\tilde{\ib{B}}$, the full matrix differential equation~\eqref{eq:gtder} now decouples into three independent equations. 
This allows us to forego exponentiating a six-by-six matrix; at most we will need to exponentiate a three-by-three matrix, which can be done analytically. 

To form the correlator matrix $\mata{}$, we take the product $\tilde{\ib{B}} \left( \uu{z} \otimes \ud{z'} \otimes\uu{v} \otimes\ud{y}\otimes \uu{x} \otimes\ud{w} \right) \tilde{\ib{B}}^\mathrm{T}$ and set $\ib{w} \to \ib{z}$ and $\ib{v} \to \ib{z'}$.
To form the transition matrix, we act with the argument of the exponential in Eq.~\eqref{eq:gt} on the operator $\uu{z} \otimes\ud{z'} \otimes\uu{v}\otimes \ud{y} \otimes\uu{x} \otimes\ud{w}$, then wedge the result between the basis vectors and set $\ib{w} \to \ib{z}$ and $\ib{v} \to \ib{z'}$:
\begin{multline}
\tilde{\ib{B}}\left[ - \frac{1}{2} \int^\eta \der \eta' \int_{\ib{u}_1,\ib{u}_2} G_{\ib{u}_1,\ib{u}_2} (\eta') \ld{a}{u_\textrm{1}} \ld{a}{u_\textrm{2}} \right]
\\
\times \left( \uu{z} \otimes \ud{z'} \otimes \uu{v} \otimes \ud{y} \otimes \uu{x}\otimes  \ud{w} \right) \tilde{\ib{B}}^\mathrm{T}
 \Bigg|_\limonemod .
\end{multline}
Diagrammatically, this is equivalent to summing all possible ways of attaching one gluon line on the operator $\; \diag{tar}$ \;, using the Fierz identity to replace the gluon vertices, and finally closing the Wilson line endpoints on the left and right using the basis vector. 
For example, the element $(6,6)$ of the correlator matrix $\mata{}$ is
\begin{align}
\nonumber
\mata{}^{(6,6)} (\eta) &= \tilde{B}_6  \left( \uu{z} \otimes\ud{z'} \otimes\uu{v} \otimes\ud{y}\otimes \uu{x} \otimes\ud{w} \right) \tilde{B}_6
\\
&= \frac{1}{\nc^3} \; \diag{m6l} \diag{tar} \diag{m6r} \; .
\end{align}
We then sum all the diagrams in which a gluon is attached to this diagram so that it joins any two of the six Wilson lines on the right of the target interaction.
Using the Fierz identity in Eq.~\eqref{eq:fierz}, we may write the resulting expression in terms of diagrams with no gluons. 
After making the substitutions $\ib{w} \to \ib{z}$ and $\ib{v} \to \ib{z'}$, the result will be a linear combination of  elements of the operator matrix $\mata{} (\eta)$. From this linear combination, one can read off the elements of column $6$ of the transition matrix $\matm{}$.
The explicit expressions for the elements of $\mata{} (\eta)$  in terms of the Wilson line correlators are shown in Appendix~\ref{appendix:matA}.

Performing this procedure for each diagram in $\mata{} (\eta)$, we get the full transition matrix
\begin{equation}
\matm{} (\eta) \Big|_\limonemod =
\begin{pmatrix}
\matm{3} & 0 & 0
\\
0 & \matm{2} & 0
\\
0 & 0 & \matm{1}  
\end{pmatrix} (\eta),
\end{equation}
where the subscripts refer to the dimension of the sub-matrix. 

The first (one-dimensional) transition sub-matrix is
\begin{align}
\matm{1} (\eta) = \cf \g{x,y}'
\end{align}
and upon exponentiation, gives the parametric equation for the dipole correlator as shown in Eq.~\eqref{eq:gtdip}. 
When inverted, this equation can be used to express the two-point function $\g{x,y}$ in terms of the dipole correlator. This will be needed to evaluate the higher-point functions in terms of the dipole $\left\langle \op{2}{x,y} \right\rangle$.

The second (two-dimensional) transition sub-matrix is
\begin{align}
\matm{2} (\eta) = \frac{\nc}{4}
\begin{pmatrix}
\matm{2}^{(1,1)}
& \matm{2}^{(1,2)}
\\
\matm{2}^{(1,2)}
& \matm{2}^{(1,1)}
\end{pmatrix} (\eta),
\end{align}
where
\begin{align}
\matm{2}^{(1,1)} (\eta) &:= \g{x,z}' + \g{y,z}' - \frac{2}{\nc^2} \g{x,y}' + \g{x,z'}' + \g{y,z'}',
\\
\matm{2}^{(1,2)} (\eta) &:= \g{x,z}' + \g{y,z}' - \g{x,z'}' - \g{y,z'}'.
\end{align}
The matrix differential equation 
\begin{align}
\partial_\eta \mata{2} (\eta) 
&= - \matm{2} (\eta) \mata{2} (\eta)
\end{align}
then gives a coupled system of $2 \times 2$ differential equations, out of which 2 are linearly independent,  corresponding to the fact that the same transition matrix operates separately on each of the two columns of $\mata{2}$. 
The exponential solution for this system of equations gives the known parametrization for the four-point correlator with one repeated coordinate~\cite{Marquet:2010cf} 
\begin{multline}
\label{eq:fourpt}
\left\langle \op{2}{x,z} \op{2}{z,y} \right\rangle
= \frac{1}{\nc^2} e^{- \cf \g{x,y}}
\\
+ \frac{2 \cf}{\nc} e^{- \cf \g{x,y}}
e^{- \frac{\nc}{2} (\g{x,z} + \g{y,z} - \g{x,y})}.
\end{multline}

The third and final (three-dimensional) transition sub-matrix is 
\begin{align}
\label{eq:m3}
\matm{3} (\eta) = 
\begin{pmatrix}
\frac{\nc}{4} \Gamma_1'
& \frac{\sqrt{\nc \cd}}{4} \Gamma_2'
& 0
\\
\frac{\sqrt{\nc \cd}}{4} \Gamma_2'
& \frac{\nc}{4} \Gamma_1'
& - \frac{1}{\sqrt{2}} \Gamma_2'
\\
0
& - \frac{1}{\sqrt{2}}  \Gamma_2'
& \Gamma_0'
\end{pmatrix},
\end{align}
where
\begin{align}
\Gamma_0 &:= \cf \g{x,y} + \nc \g{z,z'},
\\
\Gamma_1 &:= \g{x,z} + \g{y,z} - \frac{2}{\nc^2} \g{x,y} + \g{x,z'} + \g{y,z'} + 2 \g{z,z'},
\\
\Gamma_2 &:= \g{x,z} - \g{y,z} - \g{x,z'} + \g{y,z'}
\end{align}
and the primes on the $\Gamma$'s in Eq.~\eqref{eq:m3} denote derivatives in $\eta$.
Exponentiating this matrix is the last step required to get expressions for the remaining six-point correlators in $\mata{3}$. 


\subsection{Exponentiating the transition matrix $\matm{3}$}

In order to obtain the six-point functions, it is necessary to 
solve the differential equation
\begin{align}
\partial_\eta \mata{3} (\eta) = - \matm{3} (\eta) \mata{3} (\eta).
\end{align}
Solving this equation is equivalent to exponentiating the matrix $\matm{3}$, as shown above in the cases of the two- and four-point functions. 
To exponentiate $\matm{3}$, we consider two different cases: $\Gamma_2' = 0$ and $\Gamma_2' \neq 0$.

When $\Gamma_2' = 0$, $\matm{3}$ in Eq.~\eqref{eq:m3} becomes diagonal and we directly obtain 
\begin{align}
\mata{3} (\eta) = 
\begin{pmatrix}
e^{- \frac{\nc}{4} \Gamma_1}
& 0
& 0
\\
0
& e^{- \frac{\nc}{4} \Gamma_1}
& 0
\\
0
& 0
& e^{- \Gamma_0}
\end{pmatrix}.
\end{align}

When $\Gamma_2' \neq 0$, the matrix elements of $\mata{3}$ are calculated by matrix-exponentiating the full $\matm{3}$ in Eq.~\eqref{eq:m3}, giving
\begin{widetext}
\begin{align}
\label{eq:a3}
\mata{3} (\eta) = 
\sum_{i=1}^3 e^{z_i/4}\begin{pmatrix}
 \frac{a_{11}(z_i)}{d(z_i)}
& - \sqrt{\cd \nc} \Gamma_2  \frac{a_{12}(z_i)}{d(z_i)}
& - 2 \sqrt{2 \cd \nc} \Gamma_2^2  \frac{a_{13}(z_i)}{d(z_i)} 
\\ \\
- \sqrt{\cd \nc} \Gamma_2  \frac{a_{12}(z_i)}{d(z_i)}
&  \frac{m_{22}(z_i)}{d(z_i)}
& 2 \sqrt{2} \Gamma_2 \frac{a_{23}(z_i)}{d(z_i)}
\\ \\
- 2 \sqrt{2 \cd \nc} \Gamma_2^2  \frac{a_{13}(z_i)}{d(z_i)}
& 2 \sqrt{2} \Gamma_2  \frac{a_{23}(z_i)}{d(z_i)}
&  \frac{a_{33}(z_i)}{d(z_i)}
\end{pmatrix}.
\end{align}
Here, $z_i$ are the roots of the cubic polynomial 
\begin{align}
p(z) 
=& z^3 + 2 (2 \Gamma_0 + \nc \Gamma_1) z^2 
+ \left[ \nc \Gamma_1 (8 \Gamma_0 + \nc \Gamma_1) - (\nc^2 + 4) \Gamma_2^2 \right] z 
+ 4 \left[ \nc^2 \Gamma_0 \Gamma_1^2 - \left( \left( \nc^2 - 4 \right) \Gamma_0 + 2 \nc \Gamma_1 \right) \Gamma_2^2 \right].
\end{align}
They are
\begin{align}
z_1 &= \frac{1}{3} \left( - 2 c_1 + c_3 
+ \frac{1}{c_3} \left[ (c_1 - 6 \Gamma_0)^2 + 3 (\nc^2 + 4) \Gamma_2^2 \right] \right),
\\
z_2 &= - \frac{1}{6} \left( 16 c_1 + c_3 (1 - i \sqrt{3})
+ \frac{1}{c_3} (1 + i \sqrt{3}) \left[ (c_1 - 6 \Gamma_0)^2 + 3 (\nc^2 + 4) \Gamma_2^2 \right] \right),
\\
z_3 &= - \frac{1}{6} \left( 16 c_1 + c_3 (1 + i \sqrt{3})
+ \frac{1}{c_3} (1 - i \sqrt{3}) \left[ (c_1 - 6 \Gamma_0)^2 + 3 (\nc^2 + 4) \Gamma_2^2 \right] \right),
\end{align}
\end{widetext}
where
\begin{align}
c_1 =& 2 \Gamma_0 + \nc \Gamma_1,
\\
c_2 =& \left( 2 c_1 \left[ c_1^2 - 9 (\nc^2 - 8) \Gamma_2^2 \right] \right)^2 
- 4 \left[ c_1^2 + 3 (\nc^2 + 4) \Gamma_2^2 \right]^3,
\\
c_3 =& \sqrt[3]{\frac{\sqrt{c_2}}{2} + c_1 \left[ c_1^2 - 9 (\nc^2 - 8) \Gamma_2^2 \right]}.
\end{align}
Notice that $c_3$ may be complex.
The functions of the roots that appear in Eq.~\eqref{eq:a3} are
\begin{align}
a_{11} (z) &= 4 \nc \Gamma_0 \Gamma_1 - 8 \Gamma_2^2 + (2 \Gamma_0 + c_1) z + z^2,
\\
a_{12} (z) &= 4 \Gamma_0 + z,
\\
a_{13} (z) &= 1,
\\
a_{22} (z) &= 4 \nc \Gamma_0 \Gamma_1 + (2 \Gamma_0 + c_1) z + z^2,
\\
a_{23} (z) &= \nc \Gamma_1 + z,
\\
a_{33} (z) &= \nc^2 \Gamma_1^2 - (\nc^2 - 4) \Gamma_2^2 + 2 \nc \Gamma_1 z + z^2
\\
d (z) &= 3 z^2 + 4 c_1 z + \nc \Gamma_1 (8 \Gamma_0 + \nc \Gamma_1) - (\nc^2 + 4) \Gamma_2^2. 
\end{align}
Despite the fact that some of these expressions are complex, the final expressions for each element in $\mata{3}$ are in fact real, as they should be. 


\subsection{Extracting six-point correlators needed for NLO BK}

Eq.~\eqref{eq:a3} gives the analytical expressions for the correlators formed using basis $\tilde{B}$, solely in terms of the parameter $\g{}{}$ (which can be used to relate these expressions to the dipole via Eq.~\eqref{eq:gtdip}).
For example,
\begin{multline}
\mata{3}^{(3,3)} (\eta)
= \sum_{i=1}^3 e^{z_i/4} \frac{a_{33}(z_i)}{d(z_i)}
\\
= \frac{1}{\nc \da} \left\langle
\left[ - \; \diag{m4l} \; + \frac{1}{\nc} \; \diag{m6l} \; \right]
\; \diag{tar} \right.
\\
\left. \times \left[ - \; \diag{m4r} \; + \frac{1}{\nc} \; \diag{m6r} \; \right] \right\rangle_\limonemod.
\end{multline}
However, the two correlators required in Eqs.~\eqref{eq:d21} and \eqref{eq:d22} are 
\begin{align}
\left\langle \op{2}{x,z} \op{2}{z,z'} \op{2}{z',y} \right\rangle 
&= \frac{1}{\nc^3} \left\langle \; \diag{m1l} \diag{tar} \diag{m1r} \; \right\rangle_\limonemod,
\\
\left\langle \op{6}{x,z,z',y,z,z'} \right\rangle 
&= \frac{1}{\nc} \left\langle \; \diag{m5l} \diag{tar} \diag{m1r} \; \right\rangle_\limonemod.
\end{align}
These are not explicitly any of the elements of matrix $\mata{3}$, since $\; \diag{m1l} \;$ and $\; \diag{m5l} \;$ are not basis elements in $\tilde{\ib{B}}$.
Instead, they are linear combinations of the elements $\tilde{B}_i$ contained in $\tilde{\ib{B}}$:
\begin{multline}
\frac{1}{\sqrt{\nc^{3}}} \; \diag{m1l} \;
= \sqrt{\frac{\da \cd}{2 \nc^3}} \tilde{B}_1 - \frac{1}{\nc} \sqrt{\frac{\da}{2}} \tilde{B}_2 - \frac{\sqrt{\da}}{\nc^2} \tilde{B}_3 
\\
- \frac{\sqrt{2 \da}}{\nc^2} \tilde{B}_5 + \frac{1}{\nc^2} \tilde{B}_6,
\end{multline}
\begin{multline}
\frac{1}{\sqrt{\nc^{3}}} \; \diag{m5l} \;
= \sqrt{\frac{\da \cd}{2 \nc^3}} \tilde{B}_1 + \frac{1}{\nc} \sqrt{\frac{\da}{2}} \tilde{B}_2 - \frac{\sqrt{\da}}{\nc^2} \tilde{B}_3 
\\
- \frac{\sqrt{2 \da}}{\nc^2} \tilde{B}_5 + \frac{1}{\nc^2} \tilde{B}_6.
\end{multline}
Using these two expressions, it is simple to get the final equations for the two six-point correlators needed.
In terms of the elements of the correlator matrix $\mata{3}$ given in Eq.~\eqref{eq:a3} (see Appendix \ref{appendix:matA} for detailed expressions) they are
\begin{multline}
\label{eq:sss}
\left\langle \op{2}{x,z} \op{2}{z,z'} \op{2}{z',y} \right\rangle
= \frac{1}{\nc^2} \left\langle \op{6}{x,z,z',y,z,z'} \right\rangle 
\\
+ \frac{\da}{\nc^3} \left( - \sqrt{\cd \nc} \mata{3}^{(1,2)} + \nc \mata{3}^{(2,2)} + \sqrt{2} \mata{3}^{(2,3)} \right)
\end{multline}
and
\begin{multline}
\label{eq:s6}
\left\langle \op{6}{x,z,z',y,z,z'} \right\rangle
= - \langle \op{2}{x,y} \rangle + \langle \op{2}{x,z} \op{2}{z,y} \rangle + \langle \op{2}{x,z'} \op{2}{z',y} \rangle
\\
+ \frac{\da}{\nc} \left( \frac{\cd}{2} \mata{3}^{(1,1)} - \sqrt{\frac{2 \cd}{\nc}} \mata{3}^{(1,3)} \right.
\\ 
- \left. \frac{\nc}{2} \mata{3}^{(2,2)} 
+ \frac{1}{\nc} \mata{3}^{(3,3)} \right).
\end{multline}
Equations~\eqref{eq:sss} and \eqref{eq:s6} are the final two expressions needed to solve the NLO BK equation at finite $\nc$; they are the main analytical results of this work. 
It is now possible to express these six-point functions entirely in terms of dipole correlators using Eq.~\eqref{eq:gtdip}.
This makes it possible to write the NLO BK equation from Eq.~\eqref{eq:nlobk} solely in terms of dipole correlators. 
In such a closed form, it can be solved directly, as was done in the \lnc case in Refs.~\cite{Lappi:2015fma,Lappi:2016fmu}.

To verify the validity of Eqs.~\eqref{eq:sss} and \eqref{eq:s6}, we perform two checks.
Firstly, the Gaussian approximation has the built-in property that it should be consistent in color algebra. 
This means that taking any coincidence limit in which coordinates are made equal in Eqs.~\eqref{eq:sss} and \eqref{eq:s6}, should reduce them to the relevant expressions for the lower-point functions Eqs.~\eqref{eq:gtdip} and \eqref{eq:fourpt}. 
For example, setting $\ib{z} \to \ib{x}$ and $\ib{z'} \to \ib{y}$ in Eq.~\eqref{eq:sss}, we reproduce the equation for the dipole~\eqref{eq:gtdip}, as expected.  

Secondly, when Eq.~\eqref{eq:sss} is taken in the dilute limit, where the Wilson lines are expanded as 
\begin{align}
\uu{x} 
&= e^{- \lambda_a (\ib{x}) t^a} 
\\
&= 1 - \lambda_a (\ib{x}) t^a + \mathcal{O}(\lambda^2),
\qquad \lambda_a (\ib{x}) \in \mathbb{R},
\end{align}
Eq.~\eqref{eq:sss} should be the same up to order $\lambda^2$ as the parametric equation for the \lnc counterpart operator. 
In the case of correlator $\left\langle \op{2}{x,z} \op{2}{z,z'} \op{2}{z',y} \right\rangle$, the \lnc result is just the factorized product of dipole correlators 
\begin{align}
\left\langle \op{2}{x,z} \right\rangle \left\langle \op{2}{z,z'} \right\rangle \left\langle \op{2}{z',y} \right\rangle
= e^{- \cf (\g{x,z} + \g{z,z'} + \g{z',y})}.
\end{align}
After some algebra, Eq.~\eqref{eq:sss} can be shown to give the same result up to order $\g{}$.

We note that in Ref.~\cite{Dusling:2017aot}, correlators of up to eight Wilson lines have been calculated at finite $\nc$. The difference between that work and ours is that the authors there are solving the system for a general configuration of coordinates, where it is difficult to find a  basis such that the transition matrix would become block diagonal. Consequently, an analytical approach as presented in this paper is not possible.
Instead, the authors  numerically exponentiate the transition matrix, which is a much more expensive computational procedure than what is needed here.

\section{Numerical results}


We now study numerically the obtained six-point correlators, Eqs.~\eqref{eq:sss} and~\eqref{eq:s6}. 
In particular, we are interested in the effects of the $1/\nc^2$ suppressed contributions included in these six-point correlators, compared to the \lnc version in Eq.~\eqref{eq:largenc-d2}, which was used previously in numerical studies of the NLO BK equation.
We will first study these operators in a specific coordinate configuration (with the GBW parametrization for the dipole). We will then integrate the operators over the gluon coordinates $\ib{z},\ib{z'}$ and study the BK evolution starting from an MV model initial condition. 


\subsection{Correlators in a line configuration of coordinates}
\label{sec:line}

\begin{figure}[tb]
\centering
\includegraphics[height=0.13\textwidth]{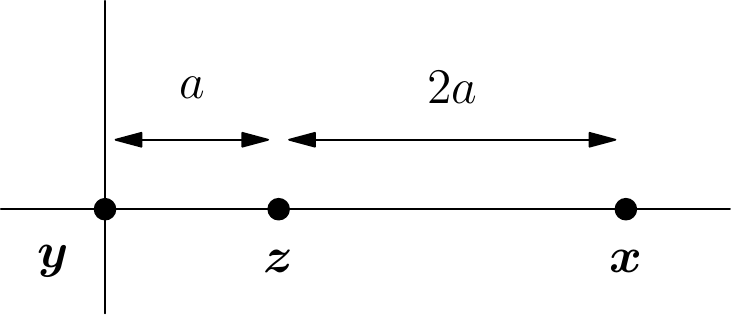}
\caption{Coordinates in the LO-like operators of the BK equation placed in a line configuration as a function of some value $a$. }
\label{fig:linelo}
\end{figure}

As a baseline for comparison of the six-point correlators in the NLO BK integrand, we consider first the four-point correlator $\left\langle \op{2}{x,z} \op{2}{z,y} \right\rangle$ that appears in the LO BK equation. We compare the full  \fnc result to its \lnc limit $\left\langle \op{2}{x,z} \right\rangle \left\langle \op{2}{z,y} \right\rangle$. The \fnc correlator is evaluated by applying Eq.~\eqref{eq:fourpt}. For the dipole operator $\left\langle \op{2}{} \right\rangle$, we use the GBW form given in Eq.~\eqref{eq:gbw}.
We consider the following, specific but not atypical, configuration of coordinates: $\ib{x}, \ib{y}$ and $\ib{z}$ in a line along the horizontal axis of transverse coordinate space, as shown in Fig.~\ref{fig:linelo}.
The distance between points $\ib{y}$ and $\ib{z}$ is denoted by $a$,  the distance between points $\ib{x}$ and $\ib{z}$ is $2a$, and  the distance between $\ib{x}$ and $\ib{y}$ is $3a$. We have confirmed that the rough relative magnitude of the \fnc effects of the results shown in this subsection are not specific to the actual chosen geometric configuration.
To show the results as a function of the dimensionless distance scale $aQ_s$, we define the saturation scale $Q_s$ as
\begin{equation}
\label{eq:qs}
\langle \op{2}{x,y} \rangle_{(\ib{x}-\ib{y})^2=2/Q_s^2} = e^{-1/2}.
\end{equation}

\begin{figure}[tb]
\centering
\includegraphics[width=0.5\textwidth]{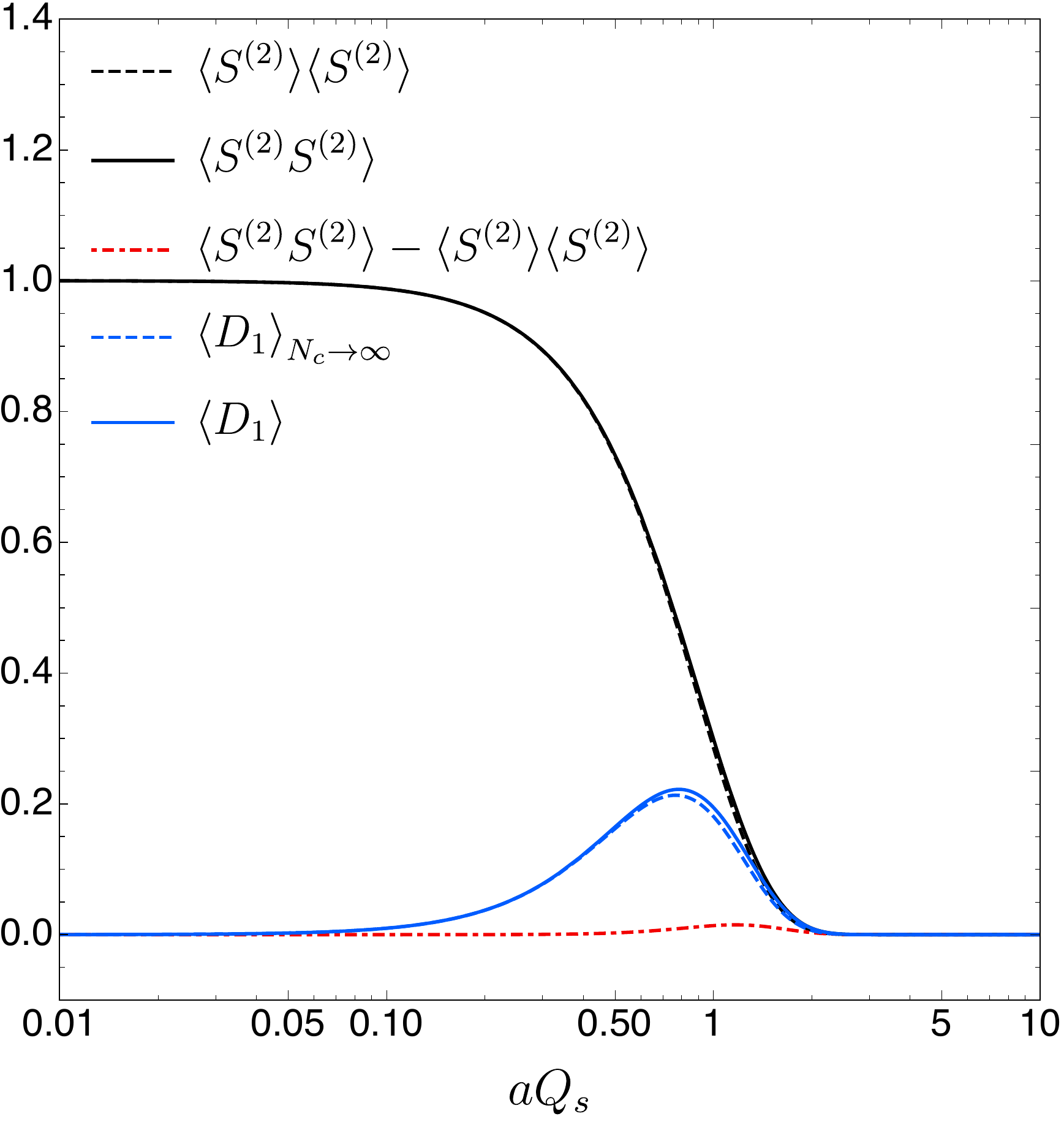}
\caption{Correlators in the LO-like piece of the BK equation~\eqref{eq:nlobk}, in the line configuration of coordinates as shown in Fig.~\ref{fig:linelo}.} 
\label{fig:lo}
\end{figure}

In Fig.~\ref{fig:lo}, the four-point correlator is shown both at finite $\nc$ and at large $\nc$ as a function of the distance $aQ_s$. 
Additionally, the magnitude of the \fnc correction is shown as a difference between the \fnc and \lnc results, denoted by $\langle \op{2}{} \op{2}{} \rangle - \langle \op{2}{} \rangle \langle \op{2}{} \rangle$.
The \fnc correction  to the four-point correlator $ \left\langle \op{2}{} \op{2}{} \right\rangle$ is found to be negligible.  At typical $a Q_s=1$, the relative \fnc correction $(\langle \op{2}{} \op{2}{} \rangle - \langle \op{2}{} \rangle \langle \op{2}{} \rangle)/(\langle \op{2}{} \rangle \langle \op{2}{} \rangle)$ is approximately $5\%$. 
The relative correction becomes more important at large $aQ_s$, in the region which gives only a negligible contribution to the BK evolution. We will return to the discussion of the \fnc corrections at $aQ_s \gtrsim 1$ later, when evaluating the six-point functions.
Also shown in Fig.~\ref{fig:lo} is the full LO-like operator factor $D_1$ (see Eq.~\eqref{eq:d1}) from the BK equation, both at large and finite $\nc$. The difference between the \fnc and \lnc results is the same as the difference for the four-point correlators. The fact that the \fnc corrections are smaller than $\sim 1/\nc^2$ is not surprising, as these corrections to the LO BK equation are known to be small~\cite{Kovchegov:2008mk}.

\begin{figure}[tb]
\centering
\includegraphics[height=0.13\textwidth]{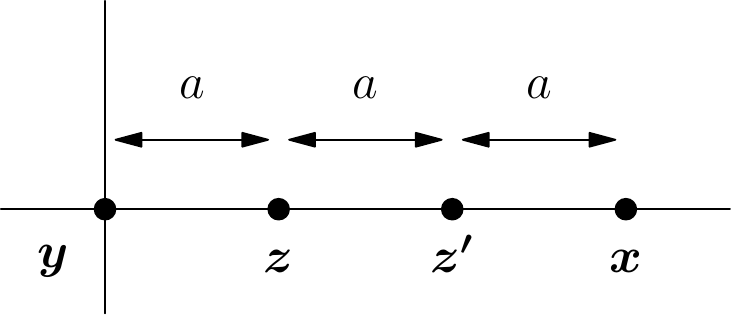}
\caption{Coordinates in the NLO-like operators of the BK equation placed in a line configuration as a function of some value $a$.} 
\label{fig:linenlo}
\end{figure}

\begin{figure}[tb]
\centering
\includegraphics[width=0.5\textwidth]{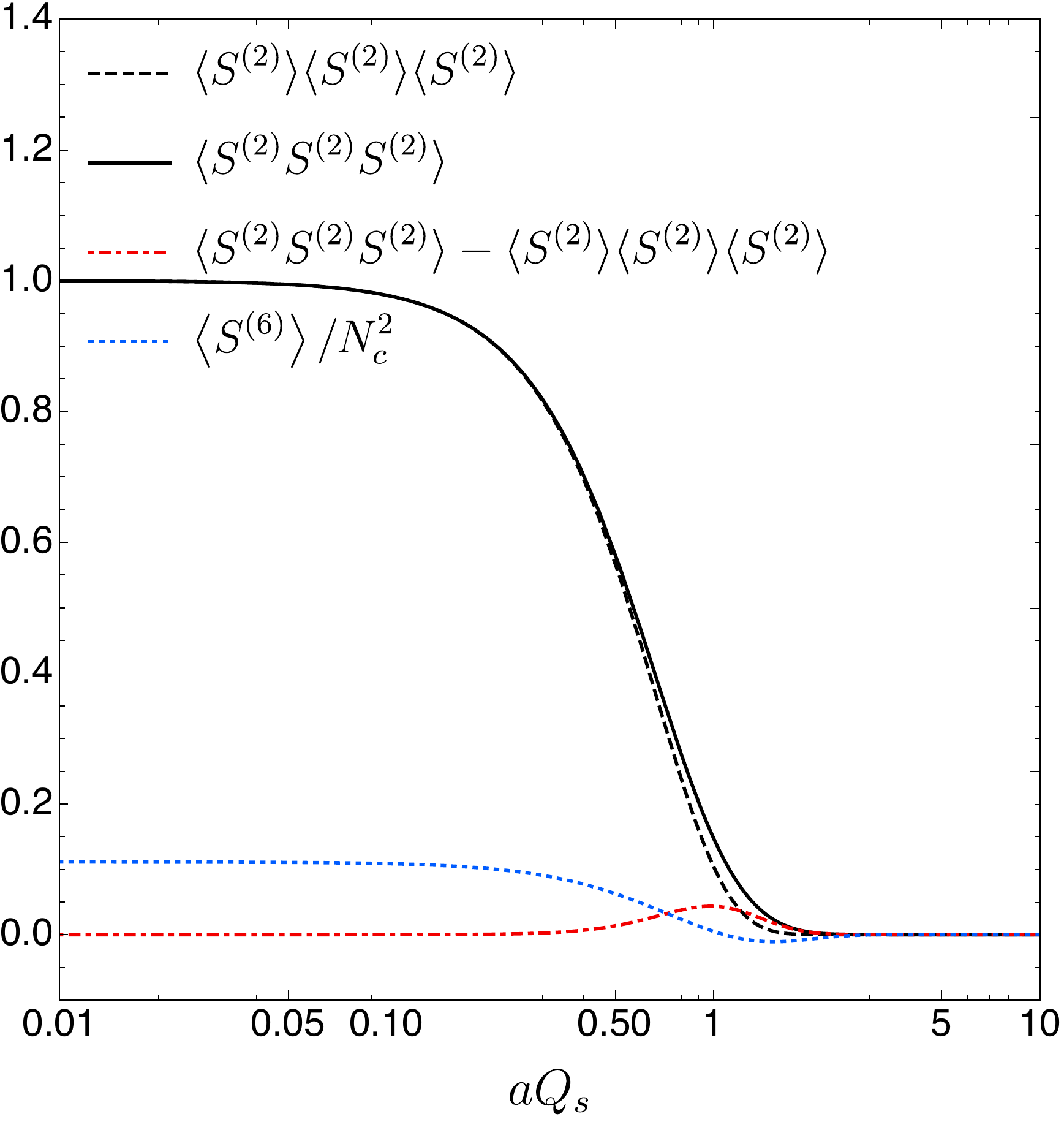}
\caption{Six-point correlators present in the NLO BK equation~\eqref{eq:nlobk}, in the line configuration of coordinates as shown in Fig.~\ref{fig:linenlo}.} 
\label{fig:nlo}
\end{figure}

Next, we choose for the four coordinates present in the NLO-like operators in the BK equation a similar line configuration, shown in Fig.~\ref{fig:linenlo}. 
In Fig.~\ref{fig:nlo}, the behaviour of the operator $\left\langle \op{2}{} \op{2}{} \op{2}{} \right\rangle$ and the \lnc counterpart $\left\langle \op{2}{} \right\rangle \left\langle \op{2}{} \right\rangle \left\langle \op{2}{} \right\rangle$ are plotted as a function of $a \qs$. 
For the \fnc correlator $\left\langle \op{2}{} \op{2}{} \op{2}{} \right\rangle$, we use our analytical result Eq.~\eqref{eq:sss}.
Although the difference between the \lnc and \fnc results (also shown in the figure) is larger here compared to the four-point function shown in Fig.~\ref{fig:lo}, it is still negligible at $aQ_s \ll 1$.
On the other hand, the \fnc corrections clearly dominate in the region  $aQ_s \gtrsim 1$, the relative contribution from $1/\nc^2$ suppressed terms being approximately $40\%$ at $aQ_s=1$. 
A similar, although numerically smaller, effect was observed in the four-point function studied above.
This can be understood as follows. When $2a \gtrsim 1/Q_s$, the color fields at points $\ib{x}$ and $\ib{z}$, as well as at $\ib{y}$ and $\ib{z'}$ are uncorrelated. However, at finite $\nc$, the six-point function is also sensitive to the color field correlations between points $\ib{x}$ and $\ib{z'}$, as well as between $\ib{y}$ and $\ib{z}$ that belong to different dipoles and are thus not correlated in the \lnc limit. When $|\ib{x}-\ib{z'}|=|\ib{y}-\ib{z}| \lesssim 1/Q_s$, these correlations do not vanish and actually dominate the full six-point function.

Also shown in Fig.~\ref{fig:nlo} for comparison is the other $1/\nc^2$ suppressed six-point correlator $\frac{1}{\nc^2} \left\langle \op{6}{} \right\rangle$ present in the NLO BK integrand at \fnc (cf. Eq.~\eqref{eq:d21}). 
This is plotted using Eq.~\eqref{eq:s6}.
We can see that the contribution of the six-point function $\op{6}{}/\nc^2$ to $D_{2,2}$ is similar in magnitude as that of the \fnc corrections to the dipole cubed operator $\op{2}{} \op{2}{} \op{2}{}$. 

\begin{figure}[tb]
\centering
\includegraphics[width=0.5\textwidth]{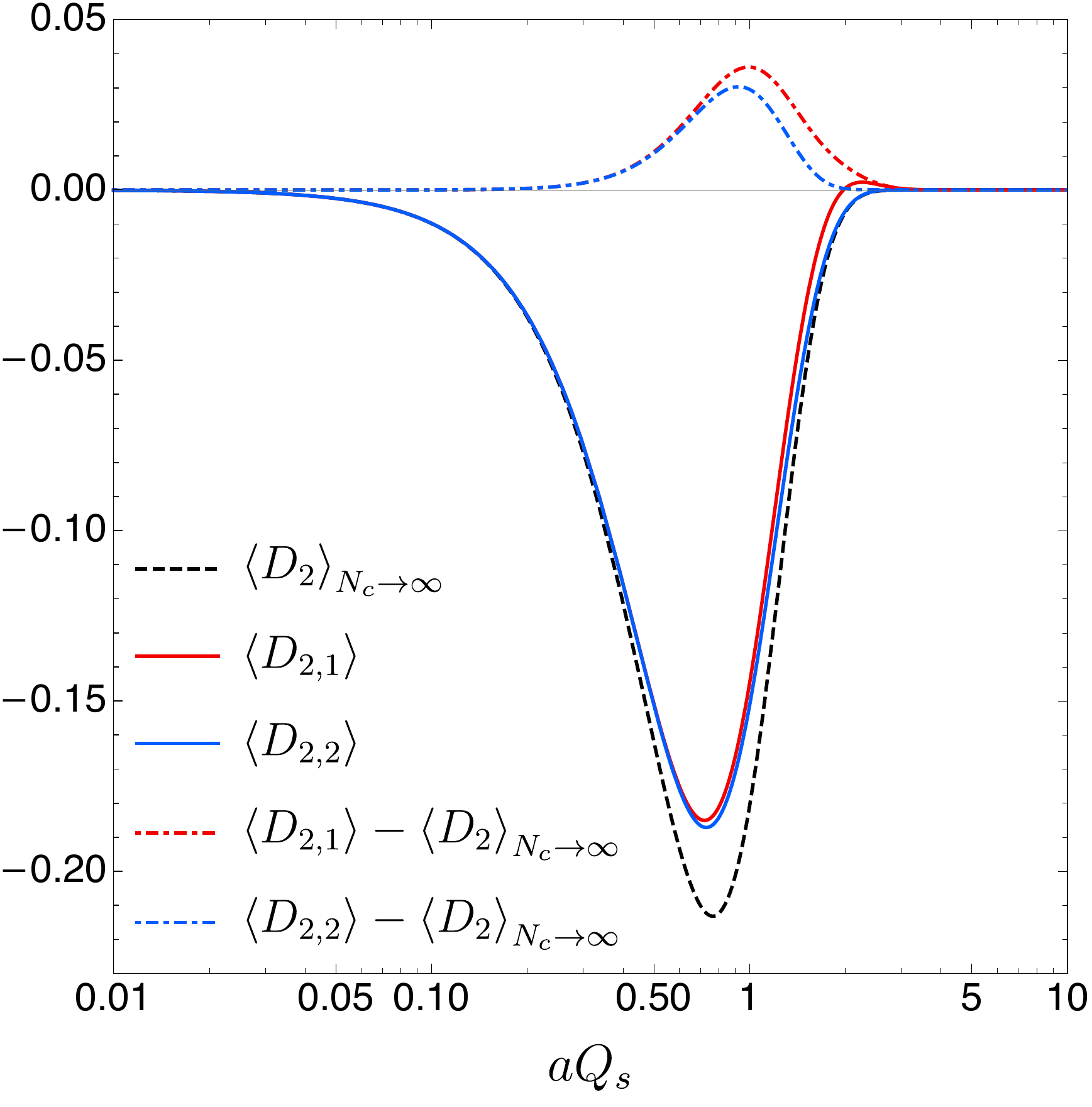}
\caption{Correlator factors $\langle D_{2,1} \rangle$ and $\langle D_{2,2} \rangle$ in the NLO-like part of the BK equation~\eqref{eq:nlobk}, in the line configuration of coordinates as shown in Fig.~\ref{fig:linenlo}. Both factors reduce to the same expression $\langle D_2 \rangle_{\textrm{Large } \nc}$ in the \lnc limit, as also shown in the figure.} 
\label{fig:d2}
\end{figure}

In Fig.~\ref{fig:d2}, we use all the above mentioned correlators to plot the NLO-like factors $\langle D_{2,1} \rangle$ and $\langle D_{2,2} \rangle$, as defined by Eqs.~\eqref{eq:d21} and \eqref{eq:d22}, respectively. 
Since both quantities reduce to the same expression in the \lnc limit, only one curve is shown for the \lnc case. 
The dashed curves show the differences, i.e. the \fnc corrections to $\langle D_{2,1} \rangle$ and $\langle D_{2,2} \rangle$.
As already seen when studying the six-point correlators, the \fnc corrections are negligible at $aQ_s\ll 1$, but become numerically important when $aQ_s \gtrsim 1$.
In comparison to the dashed curves in Fig.~\ref{fig:lo} for the LO-like case, we see that the \fnc corrections in the NLO-like case are larger. 
At $aQ_s=1$, the \fnc corrections to the operators $\langle D_{2,1}\rangle$ and $\langle D_{2,2}\rangle$ are approximately $20\%$ and $16\%$, respectively.
In comparison, the LO-like operator $\langle D_{1} \rangle$ shown in Fig.~\ref{fig:lo} has a \fnc correction of approximately $8\%$.
When considering the full NLO BK evolution, one should keep in mind that the evolution is driven by the dipole sizes $r \lesssim 1/Q_s$. As such, even though the \fnc corrections can be large at $r=1/Q_s$, the actual effect of the $1/\nc^2$ suppressed contributions to the small-$x$ evolution can be smaller. The NLO BK evolution at \fnc is studied in the next section.
 
\subsection{BK evolution at finite $\nc$}

The line configuration studied in the preceding subsection is a typical one, and the curves shown represent the expected behaviour of the correlators.
Equipped with this, we move on to studying the full BK equation using the MV model initial condition, shown in Eq.~\eqref{eq:mv}, with $Q_{s,0}^2=1\gev^2$. 
Since the \fnc corrections to the individual operators have been found to be small (except at large distances which do not significantly contribute to the BK evolution), we expect the \fnc corrections to remain small when performing integrations over gluon coordinates $\ib{z}$ and $\ib{z'}$ in Eq.~\eqref{eq:nlobk}.

\begin{figure}[tb]
\centering
\includegraphics[width=0.5\textwidth]{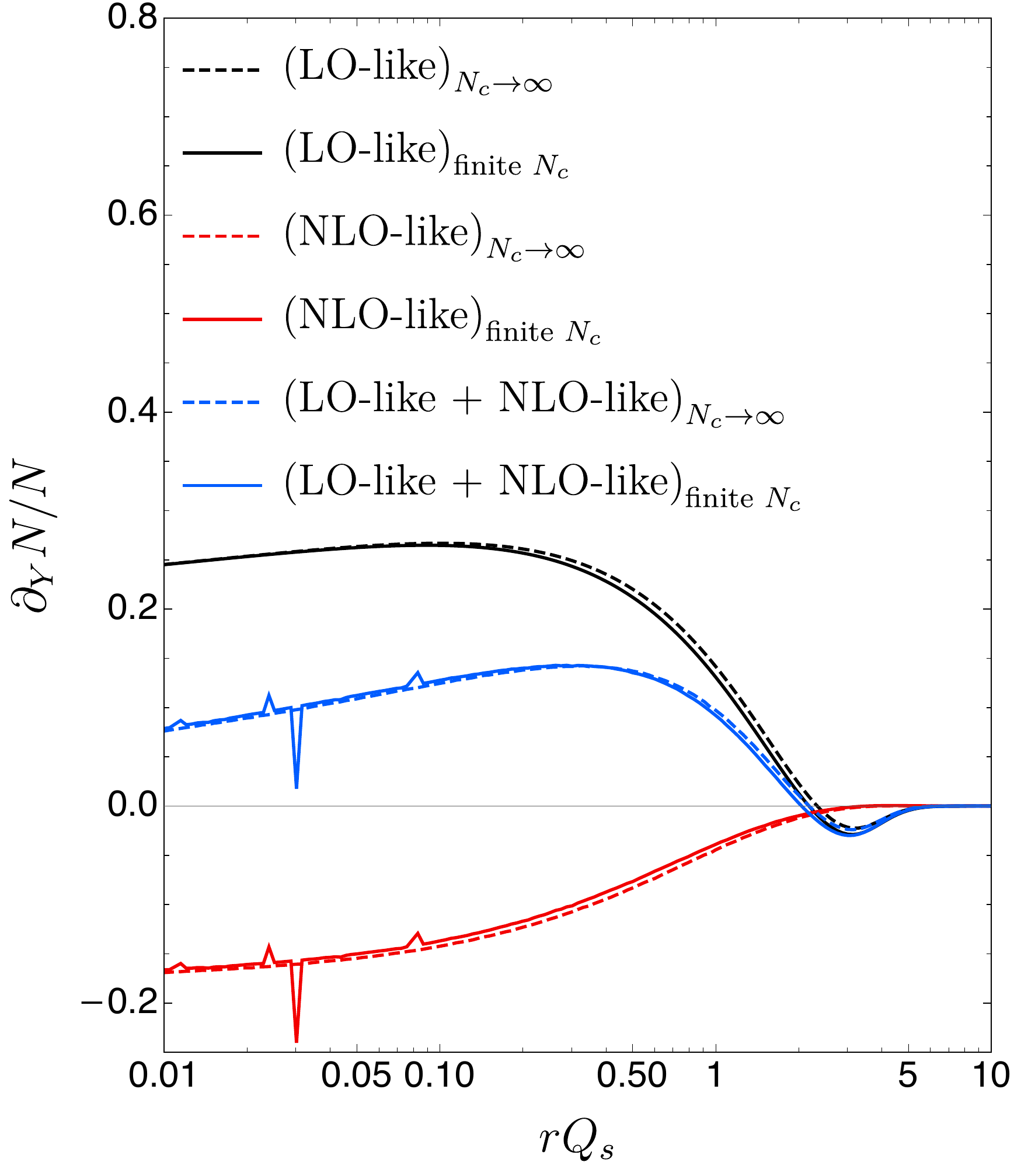}
\caption{Evolution speed of the dipole amplitude at the initial condition at \lnc and at \fnc. We show separately the contribution from the LO- and NLO-like terms (note that the LO-like contribution includes the order $\as^2$ contribution included in $K_1^\text{fin}$).}
\label{fig:dndy}
\end{figure}

In Fig.~\ref{fig:dndy}, we show the relative evolution speed $\frac{1}{N} \partial_Y N$, where the dipole amplitude $N=N_\ib{x,y}$ is defined as  $N_\ib{x,y} = \left\langle 1 - \op{2}{x,y} \right\rangle$.
This is obtained by integrating the full right side of Eq.~\eqref{eq:nlobk}, first using the \lnc expressions for the correlators, then again using the \fnc expressions. 
These are shown separately for the LO-like contribution (only the term containing $D_1$ in the integrand in Eq.~\eqref{eq:nlobk}) and the NLO-like contribution (only the terms containing $D_{2,1}$ and $D_{2,2}$). 
As expected from the line configuration studies, we see that the \fnc corrections for the NLO-like terms are slightly larger, but of the same order of magnitude as the \fnc corrections for the LO-like terms. The \fnc corrections vanish when the parent dipole size $r$ is small, and are most important at $rQ_s\sim 1$, as expected from the line configuration analysis presented above.

\begin{figure}[tb]
\centering
\includegraphics[width=0.5\textwidth]{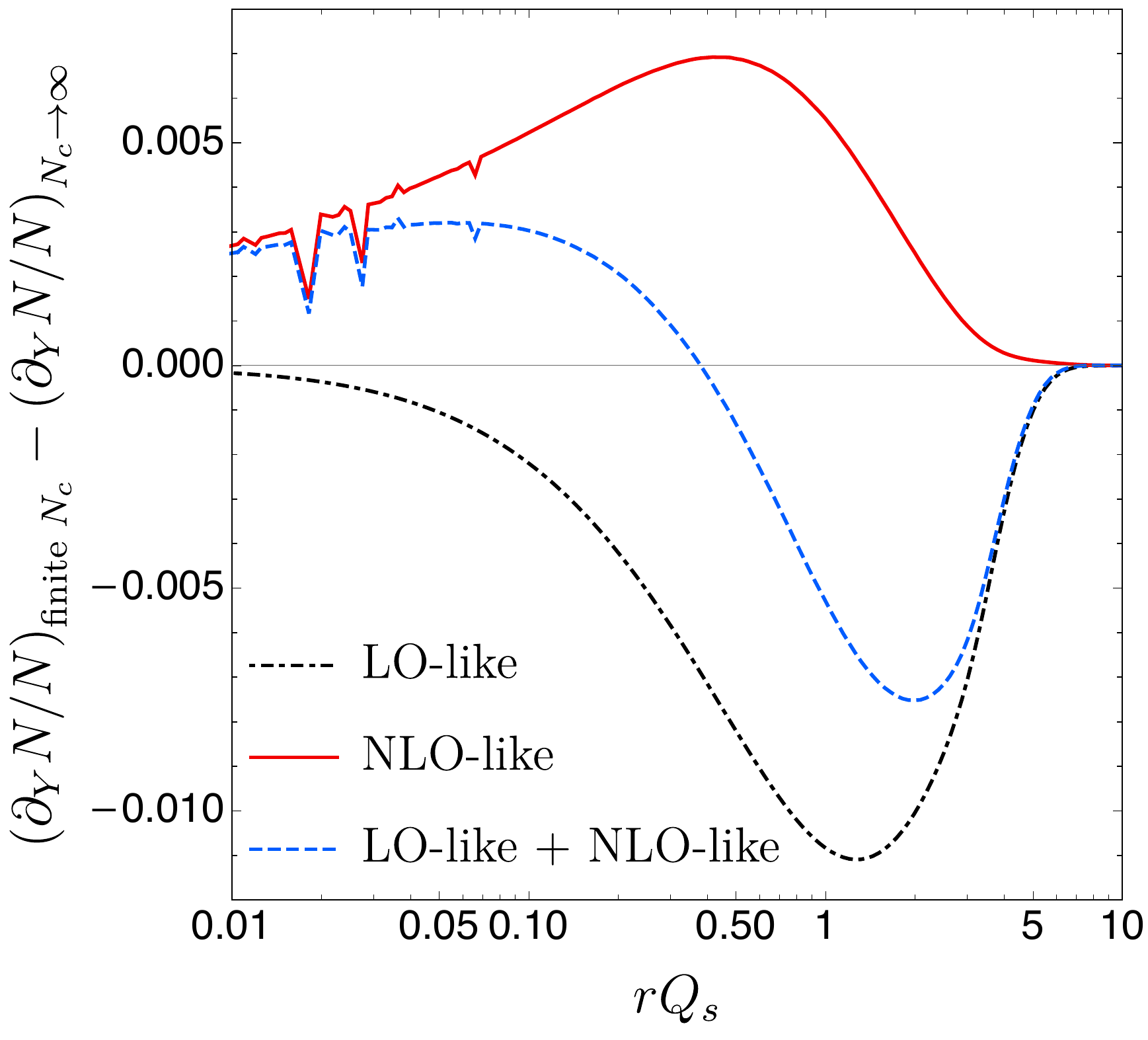}
\caption{Difference of the evolution speeds at finite $\nc$ and at large $\nc$, shown separately for the LO-like, NLO-like and total (LO-like $+$ NLO-like) contributions.  }
\label{fig:diff}
\end{figure}

In Fig.~\ref{fig:diff}, we plot the difference between the \lnc and \fnc cases, separately for the LO-like and NLO-like terms. 
This shows more clearly that the difference for the NLO-like terms is of the same order of magnitude as for the LO-like terms. 
We also note that the difference has the opposite sign in the LO-like and the NLO-like terms. Consequently, a part of the difference cancels in the total evolution speed.
At $rQ_s=1$, the relative \fnc correction is approximately $8\%$ in the LO-like contribution and $13\%$ in the NLO-like contribution.
The relative magnitude of the total $1/\nc^2$ suppressed contribution is $5\%$, which is somewhat smaller than the expected correction of $1/\nc^2 \sim 10\%$. 

\begin{figure}[tb]
\centering
\includegraphics[width=0.5\textwidth]{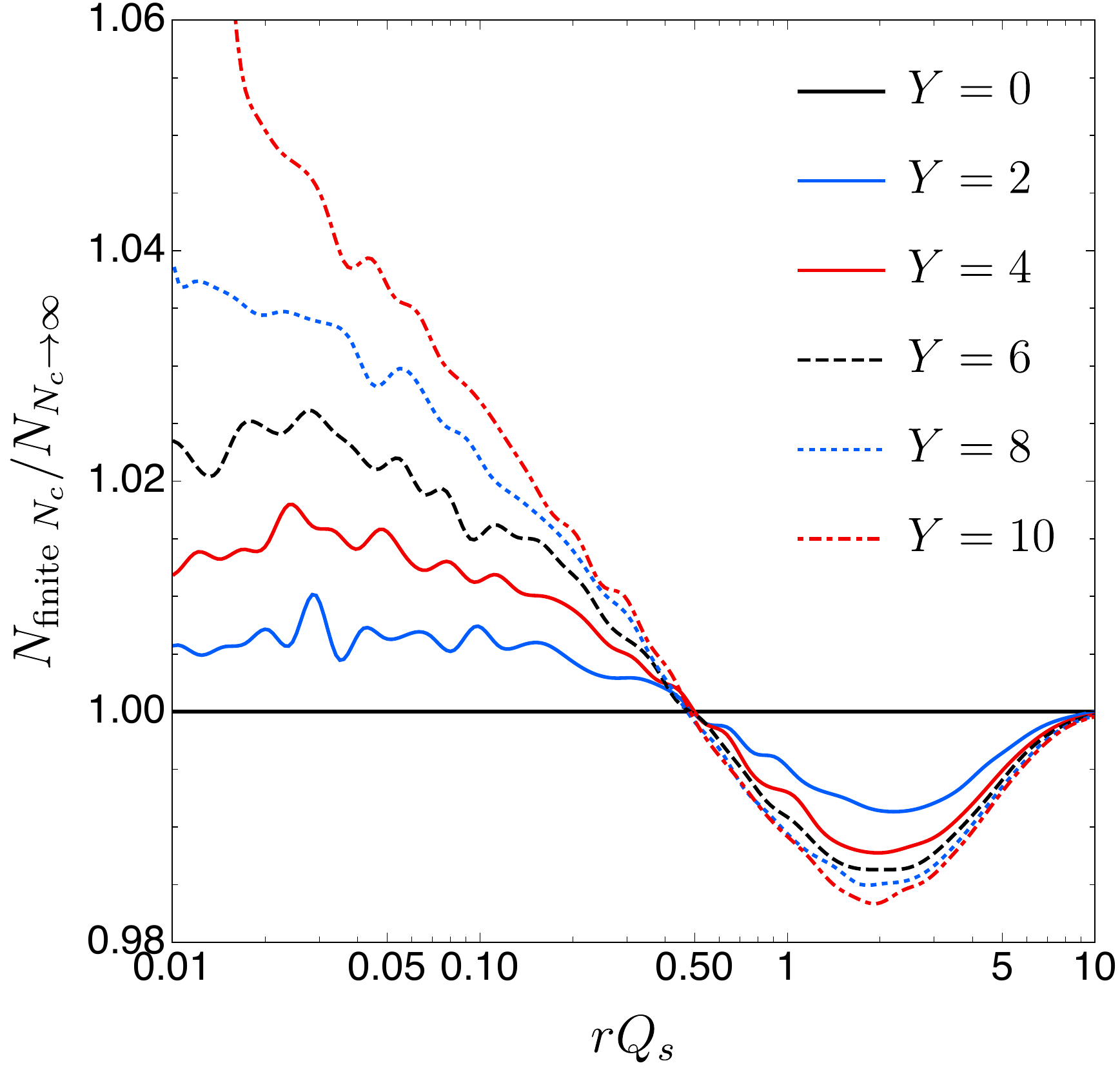}
\caption{Evolution for the ratio of the dipole amplitudes obtained by performing the \fnc and \lnc evolutions with the same initial condition. 
}
\label{fig:evol}
\end{figure}

Finally, the last thing left to study is to move beyond the initial condition and determine how the \fnc corrections behave under the NLO BK evolution. In Fig.~\ref{fig:evol}, we show the ratio of the dipole amplitudes $N$ obtained by solving the full NLO BK equation at finite $\nc$ to that at large $\nc$ \!.
At $r\gtrsim 1/Q_s$, when the details of the initial condition are lost and one enters the geometric scaling region, the difference between the \lnc and \fnc cases evolves only very slowly. At small dipoles, the ratio grows approximately linearly in $Y$. The fact that the total \fnc correction is positive at small dipole sizes and negative at large dipoles, as seen in Fig.~\ref{fig:diff}, is found to hold also asymptotically after many units of rapidity evolution.  

\begin{figure}[tb]
\centering
\includegraphics[width=0.5\textwidth]{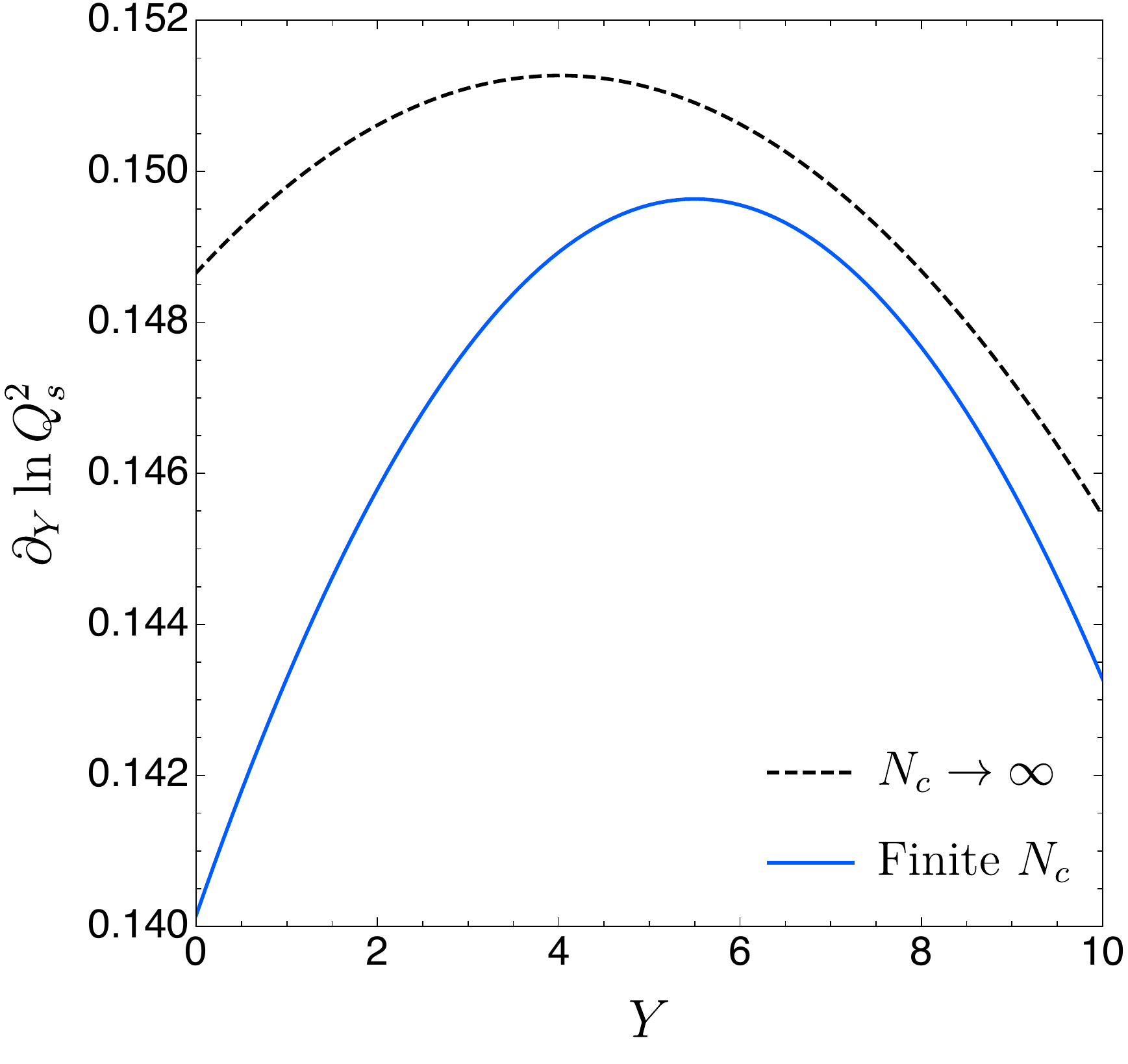}
\caption{Evolution speed of the saturation scale $Q_s^2$ as a function of rapidity at large $\nc$ and at finite $\nc$. }
\label{fig:qsevol}
\end{figure}

The evolution speed of the saturation scale,  $\partial_Y \ln Q_s^2$, 
is shown in Fig.~\ref{fig:qsevol}. Similarly to what is seen in the dipole amplitude plot in Figure~\ref{fig:evol}, we see from this figure that the \fnc corrections are more important at the initial condition, slowing down the evolution of $Q_s^2$ by approximately $5\%$. Later in the evolution, where the solution approaches the asymptotic shape of the BK evolved dipole, the difference becomes smaller - of the order of $1\%$. Consequently, even at the initial condition (and especially when the details of the initial condition are lost) the \fnc corrections to the evolution speed of $\qs$ are found to be significantly smaller than the naive expectation of $1/\nc^2$ at NLO.

\section{Conclusions}

In this work, we have studied the six-point correlators in the NLO BK equation using the Gaussian approximation.
This allowed us to express these higher-point correlators in terms of the dipole operator.
In using our analytical results, we have seen numerically that the overall \fnc corrections to the NLO-like part of the BK equation are somewhat smaller than what is naively expected. 
However, one needs to state the actual quantity being compared in order to quantify this correction.

When correlators are considered between Wilson lines separated by large distances relative to $1/\qs$, $1/\nc^2$ suppressed corrections may be considerable.
Despite these potentially large corrections to individual correlators, these configurations do not contribute much to the right side of the BK equation. 
Therefore, we find a somewhat smaller, although still significant, effect on the shape of the dipole amplitude as a function of $r$.
The \fnc corrections are watered down further when one considers the evolution speed of $\qs$ as a function of rapidity, especially once the evolution settles towards its asymptotic form away from the initial condition.
In general, \fnc corrections need to be considered carefully when evaluating the NLO BK equation, since they may have a non-negligible effect at the required accuracy.


\section*{Acknowledgements}

This work was supported by the Academy of Finland, projects 314764 (H. M.) and 321840 (T. L.), and by the National Research Foundation of South Africa (A. R.).  A. R. and T. L. are supported by the European Research Council (ERC) under the European Unions Horizon 2020 research and innovation programme (grant agreement No. ERC-2015-CoG-681707). The content of this article does not reflect the official opinion of the European Union and responsibility for the information and views expressed therein lies entirely with the authors. Computing resources from CSC -- IT Center for Science in Espoo, Finland and from the Finnish Grid and Cloud Infrastructure (persistent identifier \texttt{urn:nbn:fi:research-infras-2016072533}) were used in this work.


\begin{appendices}

\section{Correlator Matrix $\mata{} (\eta)$}
\label{appendix:matA}

We give here the explicit expressions for the operators contained in the correlator matrix $\mata{} (\eta)$. 
Since the transition matrix $\matm{}$ block-diagonalizes in basis $\tilde{\ib{B}}$, as explained in Section~\ref{sec:matm}, we are only interested in the corresponding block-diagonalized matrix
\begin{align}
\mata{} (\eta) \Big|_\limonemod =
\begin{pmatrix}
\mata{3} & 0 & 0
\\
0 & \mata{2} & 0
\\
0 & 0 & \mata{1}  
\end{pmatrix} (\eta).
\end{align}
The one-dimensional sub-matrix is 
\begin{multline}
\mata{1} (\eta) 
= \left\langle \tilde{B}_6 \; \diag{tar} \; \tilde{B}_6 \right\rangle_\limonemod
\\
= \frac{1}{\nc^3} \left\langle \; \diag{m6l} \diag{tar} \diag{m6r} \; \right\rangle_\limonemod
= \left\langle \op{2}{x,y} \right\rangle.
\end{multline}

\begin{widetext}
The two-dimensional sub-matrix is 
\begin{align}
\mata{2} (\eta) 
= \left\langle 
\begin{pmatrix}
\tilde{B}_4 
\\
\tilde{B}_5
\end{pmatrix}
\; \diag{tar} \; 
\begin{pmatrix}
\tilde{B}_4 & \tilde{B}_5
\end{pmatrix}
\right\rangle_\limonemod
= \begin{pmatrix}
\mata{2}^{(1,1)} & \mata{2}^{(1,2)}
\\ 
\mata{2}^{(1,2)} & \mata{2}^{(1,1)}
\end{pmatrix} 
(\eta),
\end{align}
where
\begin{align}
\mata{2}^{(1,1)} (\eta) 
=& \; \frac{1}{2 \nc \da} \left\langle
\; \diag{m2l} \diag{tar} \diag{m2r} \; 
- 2 \; \diag{m2l} \diag{tar} \diag{m3r} \; 
+ \; \diag{m3l} \diag{tar} \diag{m3r} \; \right\rangle_\limonemod,
\\
\mathcal{A}_2^{(1,2)} (\eta) 
=& \; \frac{1}{2 \da} \left\langle
\; \diag{m2l} \diag{tar} \diag{m2r} \;
- \; \diag{m3l} \diag{tar} \diag{m3r} \; \right\rangle_\limonemod.
\end{align}

The three-dimensional sub-matrix is 
\begin{align}
\mata{3} (\eta) 
= \left\langle 
\begin{pmatrix}
\tilde{B}_1 
\\
\tilde{B}_2 
\\
\tilde{B}_3
\end{pmatrix}
\; \diag{tar} \;
\begin{pmatrix}
\tilde{B}_1 & \tilde{B}_2 & \tilde{B}_3
\end{pmatrix}
\right\rangle_\limonemod
= \begin{pmatrix}
\mata{3}^{(1,1)} & \mata{3}^{(1,2)} & \mata{3}^{(1,3)}
\\ 
\mata{3}^{(1,2)} & \mata{3}^{(2,2)} & \mata{3}^{(2,3)}
\\
\mata{3}^{(1,3)} & \mata{3}^{(2,3)} & \mata{3}^{(3,3)}
\end{pmatrix} 
(\eta),
\end{align}
where
\begin{align}
\mata{3}^{(1,1)} (\eta)
=& \; \frac{2}{\nc^2 \da \cd} \left\langle 
2 \nc \op{2}{x,y} - \nc^3 \op{2}{x,z} \op{2}{z,y} - \nc^3 \op{2}{x,z'} \op{2}{z',y}
+ \frac{\nc^2}{4} \; \diag{m1l} \diag{tar} \diag{m1r} \;
- \nc \; \diag{m1l} \diag{tar} \diag{m4r} \; 
+ \frac{\nc^2}{2} \; \diag{m1l} \diag{tar} \diag{m5r} \; \right. \nonumber
\\
& \qquad \qquad \qquad \qquad \qquad \qquad \qquad \qquad \qquad \qquad 
\left. + \; \diag{m4l} \diag{tar} \diag{m4r} \; 
- \nc \; \diag{m4l} \diag{tar} \diag{m5r} \; 
+ \frac{\nc^2}{4} \; \diag{m5l} \diag{tar} \diag{m5r} \; \right\rangle_\limonemod,
\\
\mata{3}^{(1,2)} (\eta)
=& \; \frac{1}{2 \nc \da \sqrt{\nc \cd}} \left\langle 
- \nc \; \diag{m1l} \diag{tar} \diag{m1r} \; 
+ 2 \; \diag{m1l} \diag{tar} \diag{m4r} \; 
- 2 \; \diag{m4l} \diag{tar} \diag{m5r} \; 
+ \nc \; \diag{m5l} \diag{tar} \diag{m5r} \; \right\rangle_\limonemod,
\\
\mata{3}^{(1,3)} (\eta)
=& \; \frac{1}{\da \sqrt{2 \cd \nc}} \left\langle
- \; \diag{m1l} \diag{tar} \diag{m4r} \; 
+ \frac{2}{\nc} \; \diag{m4l} \diag{tar} \diag{m4r} \; 
- \; \diag{m4l} \diag{tar} \diag{m5r} \; \right\rangle_\limonemod,
\\
\mata{3}^{(2,2)} (\eta)
=& \; \frac{1}{2 \nc \da} \left\langle
\; \diag{m1l} \diag{tar} \diag{m1r} \; 
- 2 \; \diag{m1l} \diag{tar} \diag{m5r} \; 
+ \; \diag{m5l} \diag{tar} \diag{m5r} \; \right\rangle_\limonemod,
\\
\mata{3}^{(2,3)} (\eta)
=& \; \frac{1}{\sqrt{2} \nc \da} \left\langle
\; \diag{m1l} \diag{tar} \diag{m4r} \; 
- \; \diag{m4l} \diag{tar} \diag{m5r} \; \right\rangle_\limonemod,
\\
\mata{3}^{(3,3)} (\eta)
=& \; \frac{1}{\nc \da} \left\langle
- \nc \op{2}{x,y} 
+ \; \diag{m4l} \diag{tar} \diag{m4r} \; \right\rangle_\limonemod.
\end{align}
\end{widetext}

\end{appendices}


\bibliographystyle{JHEP-2modlong.bst}
\bibliography{refs}


\end{document}